\documentclass[a4paper,10pt,twocolumn]{article}
\pdfoutput=1
\usepackage{graphicx}
\usepackage{grffile}
\usepackage{xcolor}
\usepackage{url}
\usepackage[pdftex,colorlinks=true,linkcolor=blue,citecolor=blue,urlcolor=blue]{hyperref}
\usepackage[expproduct=cdot]{siunitx} 
\usepackage{enumitem} 

\usepackage{relsize} 
\usepackage{amsmath}
\usepackage{amssymb}
\usepackage{bbm} 
\usepackage[utopia,greekuppercase=italicized]{mathdesign}\edef\partial{\mathchar\number\partial\noexpand\!} 

\usepackage[T1]{fontenc} 

\usepackage{ctable} 
\usepackage{tabularx} 

\usepackage[numbers,compress]{natbib}

\definecolor{royalblue4}{HTML}{27408B}
\definecolor{red4}{HTML}{8B0000}
\definecolor{green4}{HTML}{008b00} 

\usepackage{dblfloatfix} 
\usepackage{fixltx2e} 

\usepackage[font=footnotesize,labelfont=bf,labelsep=endash,margin=0pt]{caption} 
\usepackage[title,header]{appendix}

\newlength{\myleftmargin} 
\newlength{\mytopmargin} 
\newlength{\myrightmargin} 
\newlength{\mybottommargin} 
\usepackage[runin]{abstract}

\let\paragraphold\paragraph
\renewcommand*{\paragraph}[1]{\paragraphold{#1.}} 
\newcommand{\keywords}[1]{\vspace{2mm}\noindent\textbf{Key words:} #1} 
\newcommand{\pagewidetitle}[3] 
{%
    \twocolumn%
        [%
            \vskip-5mm%
            \begin{@twocolumnfalse}%
                #1%
                #2%
                \vspace{5mm}%
            \end{@twocolumnfalse}%
        ]%
        #3%
}

\newlength{\figurewidth}

\newcommand*\pbuenzli[1]{#1} 

\newcommand{\ie}{{\it i.e.}}

\newcommand{\etal}{{\it et\ al.}}

\newcommand{\der}{\mathrm{d}}
\newcommand{\p}{\partial}
\newcommand{\pd}[2]{\frac{\partial #1}{\partial #2}}
\newcommand{\tpd}[2]{\tfrac{\partial #1}{\partial #2}}
\newcommand{\td}[2]{\frac{\der #1}{\der #2}}

\newcommand{\e}{\mathrm{e}}
\newcommand{\avg}[1]{\langle #1 \rangle}
\newcommand{\sign}{\mathrm{sign}}
\renewcommand{\b}[1]{{\boldsymbol{#1}}} 

\newcommand{\yr}{\ensuremath{\text{yr}}}
\newcommand{\da}{\ensuremath{\text{day}}} 
\newcommand{\days}{\ensuremath{\text{days}}} 

\newcommand{\um}{\ensuremath{\micro\metre}}

\usepackage{relsize} 
\newcommand{\bmu}{\text{BMU}} 

\newcommand{\ob}{\text{\textsmaller{O}b}}
\newcommand{\oc}{\text{\textsmaller{O}c}}

\newcommand{\ocy}{\text{\textsmaller{O}t}}

\newcommand{\kform}{\text{$k_\text{f}$}} 
\newcommand{\kres}{\text{$k_\text{r}$}}
\newcommand{\kminer}{\text{$k_\text{m}$}}

\newcommand{\form}{\text{f}}
\newcommand{\res}{\text{r}}
\newcommand{\dburial}{\text{$D_\text{burial}$}} 

\newcommand{\etagen}{\ensuremath{{\eta_\text{f}}}}

\newcommand{\avgg}[2]{\ensuremath{{\avg{#1}_{\text{\raisebox{-0.7ex}[0ex][1.1ex]{$\!#2$}}}}}}
\newcommand{\avgV}[1]{\ensuremath{{\avgg{#1}{V}}}}
\newcommand{\avgT}[1]{\ensuremath{{\avgg{#1}{\Omega}}}}
\newcommand{\vform}{\ensuremath{{v_\text{f}}}}
\newcommand{\vres}{\ensuremath{{v_\text{r}}}}
\newcommand{\sform}{\ensuremath{{S_\text{f}}}}
\newcommand{\sres}{\ensuremath{{S_\text{r}}}}
\newcommand{\Fminer}{\ensuremath{{\mathcal{F}_\text{miner}}}}
\newcommand{\mmax}{\ensuremath{{m_\text{max}}}}
\newcommand{\bv}{\text{\textsmaller{BV}}}
\newcommand{\tv}{\text{\textsmaller{TV}}}
\newcommand{\bmd}{\text{\textsmaller{BMD}}}
\newcommand{\tmd}{\text{\textsmaller{TMD}}}

\begin{document}
\title{\bf Governing equations of tissue modelling and remodelling: \\A unified generalised description of surface and bulk balance}
\author{Pascal R Buenzli$^\ast$}

\date{\small \vspace{-2mm}School of Mathematical Sciences, Monash University, Clayton VIC \pbuenzli{3800}, Australia\\\vskip 1mm \normalsize \today \vspace*{-5mm}}

\pagewidetitle{
\maketitle
}{
\begin{abstract}
Several biological tissues undergo changes in their geometry and in their bulk material properties by modelling and remodelling processes. Modelling synthesises tissue in some regions and removes tissue in others. Remodelling overwrites old tissue material properties with newly formed, immature tissue properties. As a result, tissues are made up of different ``patches'', i.e., adjacent tissue regions of different ages and different material properties, within evolving boundaries. In this paper, generalised equations governing the spatio-temporal evolution of such tissues are developed within the continuum model. These equations take into account nonconservative, discontinuous surface mass balance due to creation and destruction of material at moving interfaces, and bulk balance due to tissue maturation. These equations make it possible to model patchy tissue states and their evolution without explicitly maintaining a record of when/where resorption and formation processes occurred. The time evolution of spatially averaged tissue properties is derived systematically by integration. These spatially-averaged equations cannot be written in closed form as they retain traces that tissue destruction is localised at tissue boundaries. 

The formalism developed in this paper is applied to bone tissues, which exhibit strong material heterogeneities due to their slow mineralisation and remodelling processes. Evolution equations are proposed in particular for osteocyte density and bone mineral density. Effective average equations for bone mineral density ($\bmd$) and tissue mineral density ($\tmd$) are derived using a mean-field approximation. The error made by this approximation when remodelling patchy tissue is investigated. \pbuenzli{The specific signatures of the time evolution of $\bmd$ or $\tmd$ during remodelling events are exhibited. These signatures may provide a way to detect remodelling events at lower, unseen spatial resolutions from microCT scans.}


\end{abstract}
}{
\renewcommand{\thefootnote}{\fnsymbol{footnote}}%
\protect\footnotetext[1]{Corresponding author. Email address: \texttt{pascal.buenzli@monash.edu}}%
\renewcommand{\thefootnote}{\arabic{footnote}}%
}

\section{Introduction}
Tissue growth, renewal, and shape adaptation are common traits to many biological tissues and biomaterials. These traits are enabled by the processes of tissue modelling (tissue generation or destruction) and tissue remodelling (renewal by coordinated destruction and regeneration). Tissue growth enables us to be born small and to grow to maturity~\cite{leigh-growth}. Tissue shape adaptation and renewal enables structural reorganisation, maturation, and self-repair, which are important factors of tissue function. For example, bone tissues adapt their shape and microstructure to mechanical loads to offer strength with minimal weight, and they repair microcracks to prevent structural damage. Muscles and tendons adapt their mass and fibre structure to the forces they transmit~\cite{gollnick-etal-1973,smith-tendon}. Extracellular matrix (ECM) modelling and remodelling helps cells to migrate~\cite{molec-biol-cell} and it give cells control over local stress fields, for example to provide stress shielding~\cite{tomasek-etal-2002}.  Modelling and remodelling are often associated with the evolution of internal or external tissue boundaries (Figure~\ref{fig:tissue}), such as in wound repair and reconstruction of damaged ECM, which proceed as self-organised wave propagations~\cite{kim-etal-2012,cumming-mcelwain-upton}. Cancer invasion breaks down normal tissues boundaries, rearranging their architecture and affecting their function. 

While some tissues are renewed in a linear fashion with creation consistently occurring in one region and removal occurring in another (e.g., nail, hair, skin), other tissues have more complex patterns of creation and removal (e.g. ECM, bone), resulting in tissue heterogeneities that reflect the history of their generation.

\pbuenzli{The evolution of tissue material properties is challenging to grasp within a single mathematical modelling framework due to tissue heterogeneities and moving boundaries.} The record of maturing tissue properties may suddenly and locally be erased and overwritten with immature tissue material, creating internal discontinuities in bulk material properties within the tissue. Ordinary differential equations (ODEs) describe the time evolution of spatially averaged tissue properties, but it is unclear how changes occurring at boundaries are reflected in such spatial averages. Partial differential equations (PDEs) describe the spatio-temporal evolution of tissue properties. However, to represent discontinuities at moving interfaces, these equations must possess singular terms. The nature of these singularities is the main topic of this paper. Mathematical and computational models typically avoid such singularities by resorting to (i) volume of fluid methods or mixture theory, which represent the evolution of continuous partial fractions that in effect smooth out boundaries; or (ii) discrete models, for which discontinuities pose no particular problem~\cite{lui-chua-leong,cowin-cardoso,vanOers-etal-2008,lowengrub-etal-2010}.

In this paper, general governing equations are proposed to describe the evolution of tissue geometry and tissue properties through bulk maturation processes and through formation and resorption processes localised at tissue boundaries. \pbuenzli{The novelty of these equations} is in accounting for nonconservative, discontinuous surface balance due to creation and destruction of quantities at moving boundaries. The discontinuities associated with \pbuenzli{tissue modifications} at boundaries are captured by singular terms, namely, surface distributions, to be understood in the sense of distribution theory~\cite{lighthill,jones-distrib,buenzli2015}. These generalised material balance equations are widely applicable and as many biochemical and transport processes \pbuenzli{as necessary can be included} for a particular application. The formalism developed in this paper is anticipated to find particularly useful applications in tissue engineering, biofabrication, and investigations of bioscaffold integration and remodelling~\cite{bidan-etal-plosone-2012,bidan-etal-materialsview-2012,guyot-etal-2015}.

Surface distributions have been introduced in the context of non-equilibrium thermodynamics, interfacial conservation equations, and Stefan problems with several different representations and degrees of rigour, which has sometimes led to confusion between authors~[\citealp{bedeaux-albano-mazur-1976,ronis-bedeaux-oppenheim-1978,albano-bedeaux-vlieger-1979,gray-lee-1977,cushman,gray1982,tao-1986}; \citealp[Sec.~8.4]{jones-distrib}]. Important properties of the surface distribution and the equivalence of these representations are shown in Appendix~\ref{appx:surface-distrib-properties}.

A main advantage of formulating governing equations of tissue modelling and remodelling over discrete models, is that these equations lend themselves to mathematical analysis. We will see that this formalism enables the systematic derivation of equations governing the temporal evolution of spatial averages of tissue properties, such as density in a representative elementary volume of tissue, \pbuenzli{as well as} total tissue volume. This derivation reveals that traces that tissue removal is localised at the tissue boundary are retained in the resulting ODEs, preventing these equations from being written in closed form. The error made by closing these equations with a mean-field approximation is investigated.

\pbuenzli{A concrete application of this formalism is developed to describe} the evolution of bone tissues under bone modelling and bone remodelling \pbuenzli{processes,} with a focus on two applications of particular interest in bone: (i) evolution equations of bone-embedded cells (osteocytes); and (ii) bone mineralisation. The example~(i) extends the model of osteocyte formation and viability introduced in Ref.~\cite{buenzli2015} by including the effect of local removal by bone resorption. This extension \pbuenzli{enables the representation of} heterogenous bone states and their evolution during bone remodelling. The example (ii) is particularly important as experimental and clinical bone scans typically provide a measure of bone mineral density, averaged over spatial regions corresponding to the scanners' resolution~\cite{briggs-perilli-etal-2010}.

\section{Material balance of local tissue properties}\label{sec:balance}
\pbuenzli{Consider} a local material property of the tissue $\eta(\b r, t)$ at position $\b r$ in space and at time $t$. The value of $\eta$ is assumed to be zero out of the tissue's spatial extent. Conceptually, the definition of $\eta(\b r,t)$ involves small representative elementary volumes within which the material property is spatially averaged. These volumes are assumed large enough to contain many molecules so that the property is well-defined, but small enough so that spatial inhomogeneities occurring at a larger scale when $\b r$ is varied are not averaged out. The continuum model formally takes the limit to zero of these elementary volumes to define $\eta(\b r, t)$ at every point $\b r$ of the continuous space~\cite{landau-lifshitz,stueckelberg-scheurer,huet}. In this limit, the molecular detail is omitted and 
properties such as $\eta(\b r, t)$ become \emph{generalised functions} governed by equations to be interpreted in the sense of distributions~\cite{evans-morriss,barenblatt,jones-distrib}.

\begin{figure}[t]
    \includegraphics[width=\figurewidth]{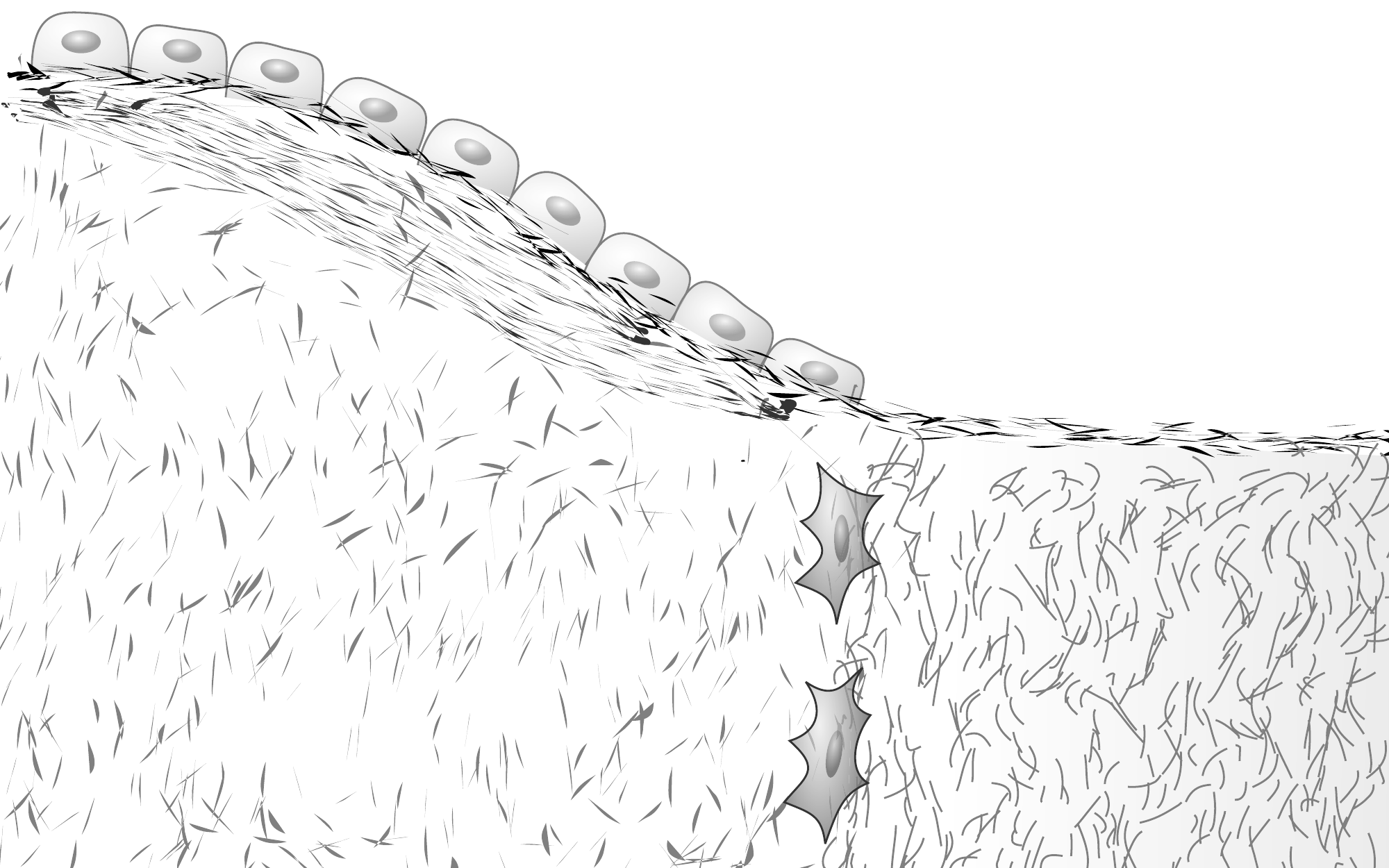}
    \caption{Cellular action on internal and external boundaries operates tissue modelling and remodelling, leading to tissue heterogeneity.}\label{fig:tissue}
\end{figure}
Tissue modelling, remodelling, and maturation modify $\eta(\b r, t)$ in several ways. Away from tissue boundaries, changes in $\eta$ are due to bulk processes such as chemical reactions and internal transport phenomena. Most of a tissue's heterogeneity is not due to such processes. It is due to the dynamic nature of tissue synthesis. Different regions of the tissue are created at different times. They have different properties $\eta$ reflecting different ages and different biological contexts at creation. Tissue heterogeneities seen in a property $\eta$ are a record of when, where, and how the tissue was synthesised or modified. We will assume here that this \pbuenzli{synthesis or modification} occurs by cellular action located at a boundary~$S(t)$, which may be an internal boundary within the tissue, or the tissue boundary (Figure~\ref{fig:tissue}). We assume quite generally that cellular action at $S(t)$ sets a new value of $\eta$ there (see Figure~\ref{fig:eta-jump}). The \pbuenzli{equation that governs} this process is given by:
\begin{align}\label{jump-eq}
\pd{}{t}\eta(\b r, t) = \Delta \eta(\b r, t)\ v(\b r, t)\ \deltaup_{S(t)}(\b r),
\end{align}
where $\Delta\eta$ is the change in $\eta$ occurring at $S(t)$ by the cellular action, $v$ is the normal velocity of $S(t)$, and $\deltaup_{S(t)}$ is the \emph{surface distribution}, formally zero everywhere except at $S(t)$, where it is infinite. This singularity indicates the discontinuous nature of $\eta$ at $S(t)$. It will be responsible for the creation of sharp internal boundaries within the tissue when the normal velocity of $S(t)$ changes sign, for example at reversals between tissue resorption and tissue formation. Mathematically, $\deltaup_S(\b r)$ is a distribution defined such that it maps any test function $\varphi(\b r)$ to the real value given by the surface integral of $\varphi$ over $S$~[\citealp{bedeaux-albano-mazur-1976,ronis-bedeaux-oppenheim-1978,albano-bedeaux-vlieger-1979}; \citealp[Sec.~8.4]{jones-distrib}]:
\begin{align}\label{def-surface-distrib}
    \deltaup_S: \varphi \mapsto \int\!\!\der\b r\ \deltaup_S\, \varphi \equiv \int_S\!\! \der\sigma\ \varphi.
\end{align}
Several properties of $\deltaup_S$ are presented in Appendix~\ref{appx:surface-distrib-properties}. The justification of Eq.~\eqref{jump-eq} is given by the following jump property:

\begin{figure}[t]    \centering\includegraphics[width=0.9\figurewidth]{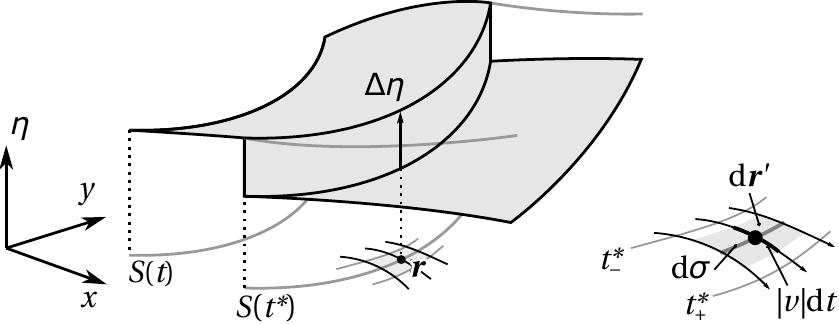}
    \caption{The tissue material property $\eta(\b r, t)$ at $\b r$ jumps by the quantity $\Delta \eta$ at the arrival time $t^\ast(\b r)$. The sign of the jump depends on the direction of propagation of the \pbuenzli{boundary}. Right: in a small neighbourhood of $\b r$, it is possible to choose a local coordinate system with components parallel to and perpendicular to $S\big(t^\ast(\b r)\big)$ such that the infinitesimal volume element $\der\b r = \der\sigma |v|\der t$ (see also explanations in the text).}\label{fig:eta-jump}
\end{figure}
\paragraph{Jump property}
{\it Let $\eta$ be governed by Eq.~\eqref{jump-eq} and let $t^\ast = t^\ast(\b r)$ be the \emph{arrival time} of the boundary $S(t)$ at point $\b r$, such that $\b r\in S(t^\ast)$ (see Figure~\ref{fig:eta-jump}). Then, the value of $\eta$ at $\b r$ is constant except at time $t^\ast$ where it jumps by the quantity $\Delta \eta$:
\begin{align}\label{jump}
    \eta(\b r, t^\ast_+) - \eta(\b r, t^\ast_-) = \sign\big(v(\b r, t^\ast)\big)\ \Delta\eta(\b r, t^\ast),
\end{align}
where $t^\ast_\pm = t^\ast \pm \epsilon$, and $\epsilon\stackrel{>}{\to} 0$.
}

\smallskip
This property is demonstrated in Appendix~\ref{appx:jump-property}. \pbuenzli{One can see from Figure~\ref{fig:eta-jump} that the sign of the jump in $\eta$ depends on the direction of propagation of the boundary, and therefore on the sign of $v$: if the boundary in Figure~\ref{fig:eta-jump} travels toward the right, the value of $\eta$ at $\b r$ increases by $\Delta\eta>0$ at the passage of the boundary; if the boundary travels toward the left, the value of $\eta$ at $\b r$ decreases by $\Delta\eta$ at the passage of the boundary. In practice, the normal velocity of cells at tissue boundaries can always be assumed positive and the sign of the jump is then solely determined by that of $\Delta\eta$: a positive sign represents formation, a negative sign represents resorption.} The jump in $\eta$ at $\b r$ only depends on the values of $v$ and $\Delta\eta$ prevailing at the arrival time $t^\ast$. At any other time than the arrival time $t^\ast$, Eq.~\eqref{jump-eq} implies that $\eta$ at $\b r$ is constant. 

The full balance of a general tissue property $\eta$ is obtained by adding to Eq.~\eqref{jump-eq} further conservative and nonconservative processes that modify $\eta$ at other times. For illustration, consider a tissue occupying a region $\Omega(t)$ in space with boundary $S(t) = \p\Omega(t)$. The tissue is assumed to change shape due to tissue formation and resorption occurring at specific regions of $S(t)$. It is also assumed to change its material properties due to maturation. The surface and bulk balance of a property~$\eta$ of this tissue can be formulated based on Eq.~\eqref{jump-eq} as follows:
\begin{itemize}
\item\emph{Tissue formation.} New tissue is deposited on $S(t)$ with a normal velocity $v=\vform>0$ and an initial material property $\etagen(\b r, t)$. Both $\vform$ and $\etagen$ are determined by the synthesis process (e.g., cell secretion). By Eq.~\eqref{jump-eq}, the rate of change in $\eta$ due to this process is $\eta_f \vform \deltaup_{S(t)}$.

\item\emph{Tissue resorption.} Existing tissue is resorbed from $S(t)$ with a normal velocity $v= - \vres< 0$ determined by the removal process (e.g., cell-driven chemical dissolution or mechanical wear). The property $\eta$ drops from its current value to zero. By Eq.~\eqref{jump-eq}, the rate of change in $\eta$ due to this process is $ - \eta_- \vres \deltaup_{S(t)}$, where $\eta_-$ is the value of $\eta$ probed at an infinitesimal inward offset of $S(t)$.

\item\emph{Tissue maturation.} After new tissue synthesis, $\eta$ evolves according to biochemical and mechanical processes specific to $\eta$, until it is removed by resorption. The rate of change in $\eta$ due to this process is assumed to be given by a maturation law $\mathcal{F}(\eta; \b r, t)$.
\end{itemize}

The evolution of $\eta$ is given by summing up these contributions:
\begin{align}\label{gov-eq}
    \pd{}{t} \eta(\b r, t)\ =\ \etagen\ \vform\,\deltaup_{S(t)}\ -\ \eta_-\, \vres\, \deltaup_{S(t)}\ +\ \mathcal{F}(\eta).
\end{align}
Some regions of $S(t)$ may undergo formation while others may undergo resorption simultaneously. \pbuenzli{Since these regions may not overlap, the} normal velocity is given everywhere by $v = \vform - \vres$, \pbuenzli{where} $\vform$ and $\vres$ correspond to the positive and negative parts of $v$. The evolution of the tissue's shape is univocally determined by~$v$~\cite{sethian,osher-fedkiw}.

The regularisation $\eta_-$ in the resorption term is necessary because $\eta$ is discontinuous at $S(t)$. It ensures that the value of $\eta$ to remove during resorption is probed at a point lying just within the tissue rather than where it jumps to 0. In the sequel, we will omit this regularisation from the notation with the convention that $\eta$ takes the value $\eta_-$ whenever it is evaluated at~$S(t)$.

The maturation law $\mathcal{F}$ represents a general bulk balance of~$\eta$, which may include nonconservative processes such as chemical reactions, and conservative processes due to transport phenomena within the tissue. 

\paragraph{Nonconservative vs conservative surface balance}
Equation~\eqref{gov-eq} is a generalised balance equation that explicitly accounts for nonconservative processes occuring at moving interfaces due to creation and destruction of material. It has similar surface terms as conservation equations in multiphase systems and Stefan problems~\cite{whitaker1973,bedeaux-albano-mazur-1976,ronis-bedeaux-oppenheim-1978,deemer-slattery,albano-bedeaux-vlieger-1979,whitaker1992,irschik,slattery-sagis-oh,tao-1986}. The main difference is that surface terms in these systems are inherently conservative. They represent jump conditions necessary to enforce mass conservation at the interface. To illustrate the difference, consider the general transport theorem that expresses the total variation of $\eta$ in an evolving domain $\Omega(t)$~\cite{whitaker1992}. Taking $\Omega(t)$ to follow the material velocity of $\eta$ such that there is no influx or outflux of $\eta$ through $\p\Omega(t)$, one has
\begin{align}\label{general-transport-theorem}
    \td{}{t}\int_{\Omega(t)}\hspace{-1em}\der\b r\ \eta = \int_{\Omega(t)}\hspace{-1em}\der\b r\ \pd{\eta}{t} + \int_{\p\Omega(t)}\hspace{-1.5em} \der\sigma\ v \eta = \int_{\Omega(t)}\hspace{-1em}\der\b r\ \mathcal{F}(\eta),
\end{align}
where the term in the right hand side represents change in $\eta$ within $\Omega(t)$ due to nonconservative phenomena such as chemical reactions.
The surface integral can be rewritten $\int_{\p\Omega(t)}\der\sigma\, v\eta = \int_{\Omega(t)}\der\b r\, \deltaup_{\p\Omega(t)} v \eta$. Because $\Omega(t)$ is an arbitrary region of the substance, $\eta$ must be governed locally by:
\begin{align}\label{general-transport-theorem-local}
    \pd{\eta}{t} = \eta\,v\,\deltaup_{\p\Omega(t)} + \mathcal{F}(\eta) = \eta_-\,\vform\,\deltaup_{\p\Omega(t)} - \eta_-\,\vres\,\deltaup_{\p\Omega(t)} + \mathcal{F}(\eta).
\end{align}
In Eq.~\eqref{general-transport-theorem-local}, the conservative balance of $\eta$ imposes the fact that the jump in $\eta$ at locations of $S(t)$ with a positive normal velocity $v=\vform$, is the value of $\eta$ at an infinitesimal inward offset of $\p\Omega(t)$, rather than an independent value $\etagen$ determined by nonconservative processes as in Eq.~\eqref{gov-eq}. Furthermore, the normal velocity $v$ of the boundary in Eqs~\eqref{general-transport-theorem}--\eqref{general-transport-theorem-local} is determined by the material velocity of the substance $\eta$, whereas in Eq.~\eqref{gov-eq}, it is determined by the independent processes of new tissue formation and resorption occuring at the interface. Naturally, both conservative and nonconservative surface balance terms may in general be present in the balance of a property.

\section{Evolution of spatially averaged tissue properties}
Many mathematical models describe the evolution of tissue properties in time only. These models implicitly assume that the property is distributed homogeneously in the tissue. Equation~\eqref{gov-eq} enables us to derive systematically the time evolution of spatial averages of patchy tissue properties, and to investigate the error made by assuming tissue homogeneity. (See Refs~\cite{whitaker1973,gray-lee-1977,cushman,whitaker1992,lasseux-etal} for volume averaging theorems in conservation equations.) Let $V$ be a fixed mesoscopic or macroscopic representative elementary volume and $\Omega(t)$ be the volume occupied by the tissue in $V$\pbuenzli{, with boundary $S(t)=\p\Omega(t)$}. The tissue volume fraction in $V$ is
\begin{align}
    f(t) \equiv \frac{\Omega(t)}{V} \leq 1.
\end{align}
(We use $V$, $\Omega(t)$, and $S(t)$ to denote both the region in space and the measures $|V|$, $|\Omega(t)|$, and $|S(t)|$ for simplicity.)
Two spatial averages of $\eta$ can be defined based on~$V$ and~$\Omega(t)$:
\begin{align}\label{avgV}
    &\avgV{\eta} \equiv \frac{1}{V}\int_V \hspace{-0.5em}\der \b r\ \eta(\b r, t),
    \\&
    \avgT{\eta} \equiv \frac{1}{\Omega(t)}\int_{\Omega(t)} \hspace{-1em}\der \b r\ \eta(\b r, t).\label{avgT}
\end{align}
The average $\avgV{\eta}$ may integrate $\eta$ over regions devoid of tissue\pbuenzli{, where $\eta=0$}. It is \pbuenzli{thus} related to $\avgT{\eta}$ through the tissue volume fraction:
\begin{align}\label{avgV-avgT}
    \avgV{\eta} = f(t)\ \avgT{\eta}.
\end{align}
Differentiating Eq.~\eqref{avgV} with respect to $t$ and using Eq.~\eqref{gov-eq} gives:
\begin{align}
    \td{\avgV{\eta}}{t} &= \frac{1}{V}\!\!\int_{S(t)}\hspace{-1em}\der\sigma\ \vform\, \etagen - \frac{1}{V}\!\!\int_{S(t)}\hspace{-1em}\der\sigma\ \vres\, \eta + \avgV{\mathcal{F}(\eta)}\notag
    \\& = \frac{S(t)}{V} \left[\avgg{\vform\,\etagen}{S} - \avgg{\vres\,\eta}{S}\right] + \avgV{\mathcal{F}(\eta)},
\label{avg1}
\end{align}
\pbuenzli{where} $\avgg{\cdot}{S}=\frac{1}{S(t)}\int_{S(t)}\der\sigma\, \cdot$ is the average value over $S(t)$. The surface density $S(t)/V$ (also called specific surface) is an important characteristic of porous media. For example in bone tissues, it is related to the propensity to remodel~\cite{martin1984,lerebours-spec-surf}. Let $\sform(t)$ and $\sres(t)$ denote the forming and resorbing surfaces of $S(t)$, i.e., the portions of $S(t)$ at which $\vform \neq 0$ and $\vres \neq 0$,  respectively. Equation~\eqref{avg1} can be rewritten as:
\begin{align}\label{avg2}
    \td{\avgV{\eta}}{t} = \frac{\sform(t)}{V}\avgg{\vform\,\etagen}{\sform} - \frac{\sres(t)}{V}\avgg{\vres\,\eta}{\sres} + \avgV{\mathcal{F}(\eta)}.
\end{align}

If $\eta$ is taken to be the indicator function $\mathbbm{1}_{\Omega(t)}$ of $\Omega(t)$, then $\avgV{\mathbbm{1}_{\Omega(t)}} = f(t)$ and Eqs~\eqref{avg1},~\eqref{avg2}, together with the balance equation of the indicator function (Eq.~\eqref{indicator-gov-eq}, Appendix~\ref{appx:indicator-gov-eq}) determine the evolution of the tissue volume fraction $f(t)$:
\begin{align}\label{f-gov-eq}
    \td{f(t)}{t} = \frac{S(t)}{V} \avgg{v}{S} = \frac{S_\form(t)}{V} \avgg{\vform}{\sform} - \frac{S_\res(t)}{V}\avgg{\vres}{\sres}.
\end{align}
Equation~\eqref{f-gov-eq} shows in particular that the tissue volume $\Omega(t)=V f(t)$ evolves according to:
\begin{align}\label{tissue-volume}
    \td{\Omega(t)}{t} = S(t)\ \avgg{v}{S},
\end{align}
i.e., tissue volume changes at a rate equal to the tissue surface area multiplied by the average normal velocity, as expected. To determine the evolution of averages defined with $\Omega(t)$ as volume referent, note that from Eq.~\eqref{avgV-avgT}:
\begin{align}
    \td{\avgT{\eta}}{t} = \frac{1}{f(t)}
\left( \td{\avgV{\eta}}{t} - \avgT{\eta}\td{f(t)}{t} \right).
\end{align}
Using Eqs~\eqref{avg2} and~\eqref{f-gov-eq}, one obtains
\begin{align}
    \td{\avgT{\eta}}{t} = \frac{1}{f(t)}
\Bigg[&\frac{S_f(t)}{V}\Big(\avgg{\vform\,\etagen}{\sform} - \avgg{\vform}{\sform}\avgT{\eta}\Big)\label{avg3}
    \\&- \frac{S_r(t)}{V}\Big(\avgg{\vres\,\eta}{\sres} - \avgg{\vres}{\sres}\avgT{\eta}\Big)\Bigg] + \avgT{\mathcal{F}(\eta)}.\notag
\end{align}

Equations~\eqref{avg2} and \eqref{avg3} show that the evolution of spatial averages of patchy tissues cannot be written in closed form even when $\mathcal{F(\eta)}$ is linear, i.e., even when $\avg{\mathcal{F}(\eta)} = \mathcal{F}(\avg{\eta})$. Indeed, due to resorption, changes in $\avgV{\eta}$ or $\avgT{\eta}$ depend on the value of $\eta$ deposited last, occurring in the factor $\avgg{\vres\,\eta}{S_r}$
, rather than on the current volume average. This hysteresis of the evolution of averages is due to the fact that tissue resorption proceeds from the tissue surface, and thus removes a value of $\eta$ that depends on when and how it was first deposited. In Section~\ref{sec:mean-field}, the error committed when closing the equations using a mean-field approximation is studied on bone mineral density. Note that if the material property does not mature \pbuenzli{($\mathcal{F}\equiv 0$)} and if a constant value $\etagen$ is generated during tissue formation, then there is no hysteresis, and trivially, $\avgT{\eta}=\etagen$, $\avgV{\eta} = \etagen f(t)$.

\begin{figure*}[!thb]
    \centering
    \begin{tabular}{ll}
    \small (a) & \small (b)\vspace*{-5mm}
    \\
      \includegraphics[width=\figurewidth]{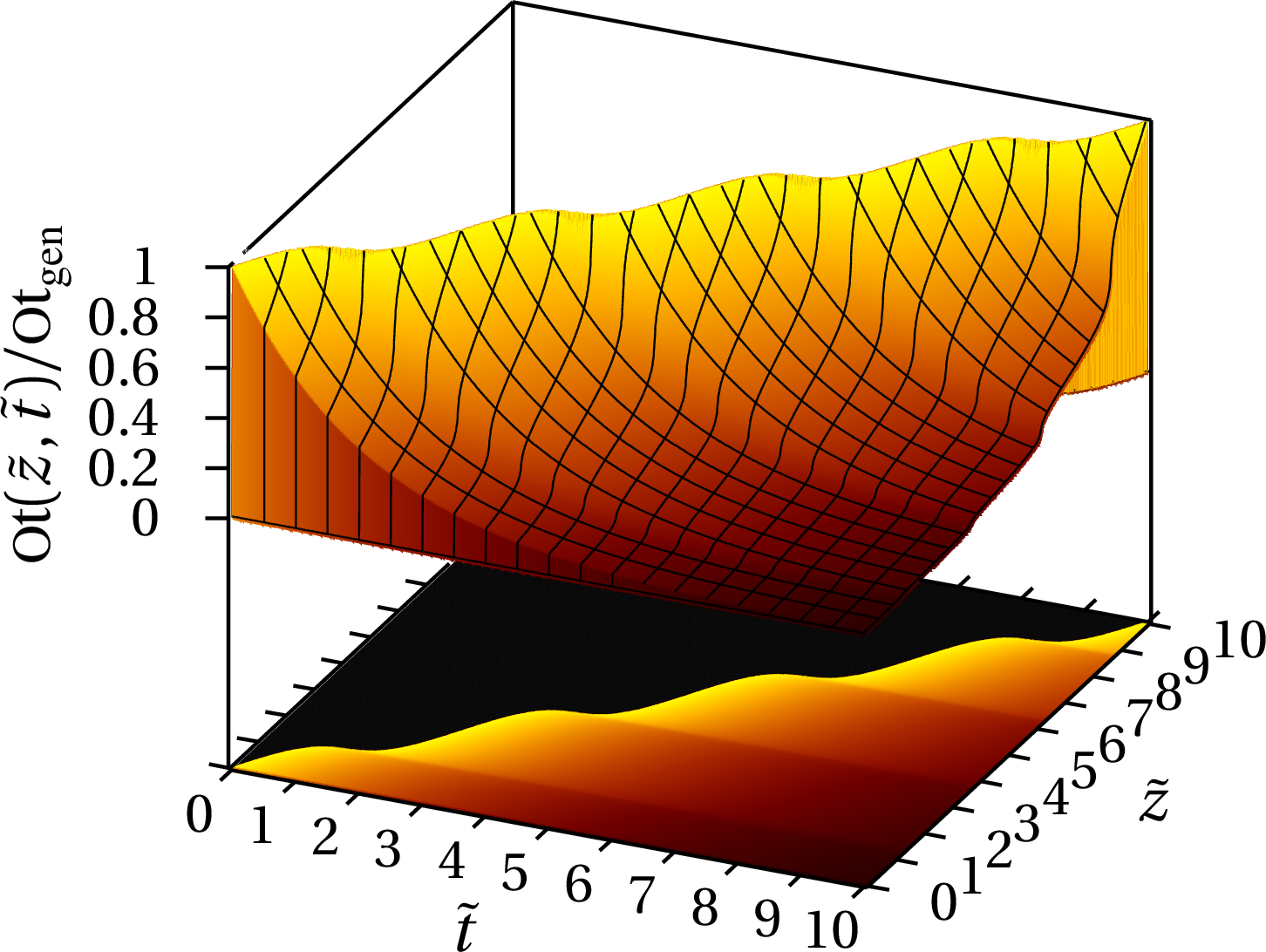}
               & \includegraphics[width=\figurewidth]{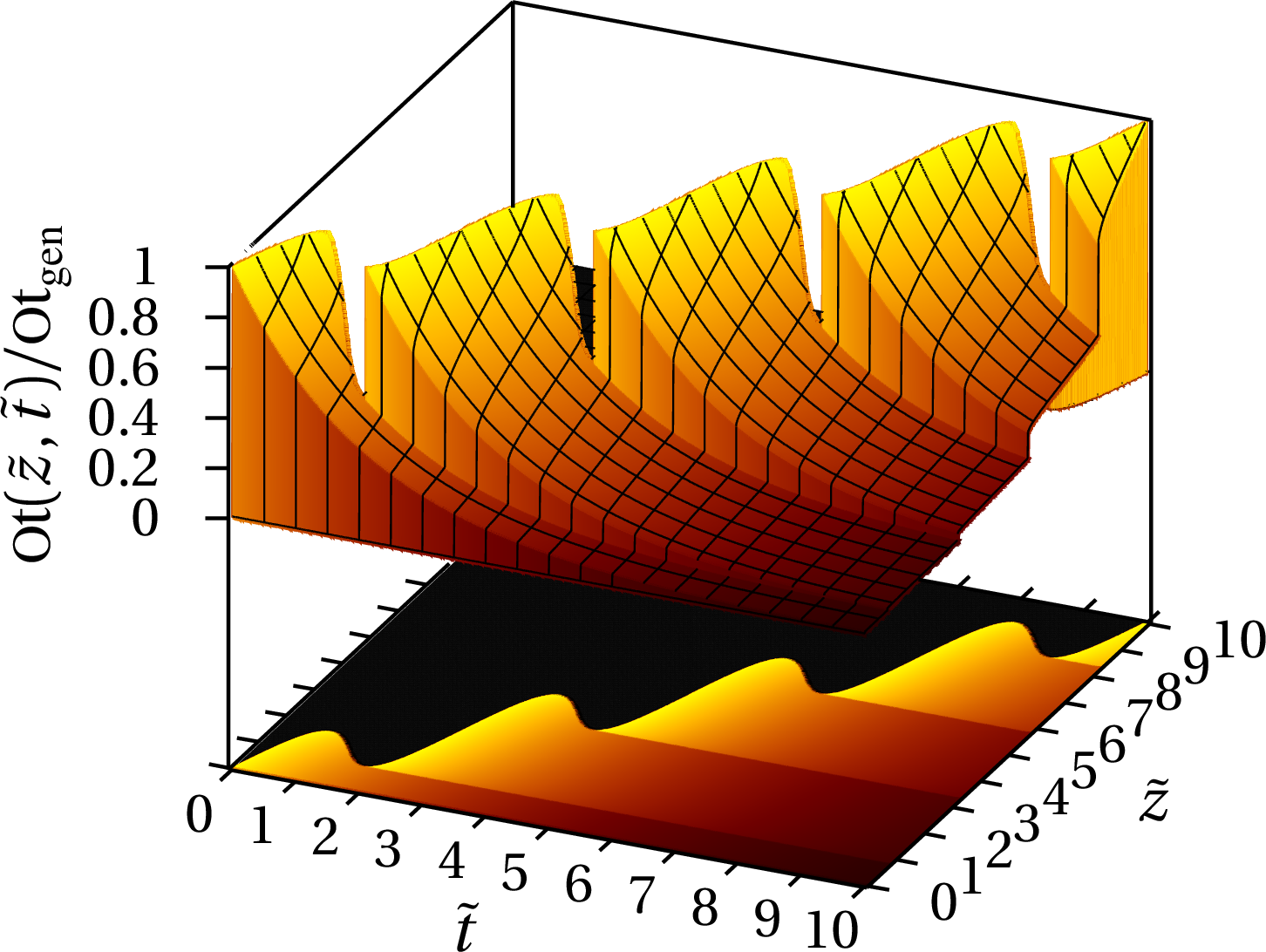}
    \end{tabular}
\caption{Spatio-temporal plot of the density of osteocytes in one spatial dimension. The cells are assumed to be generated with uniform density at the moving deposition front, and to undergo cell death at constant rate. The variables $\tilde{t}=t\,A/2$ and $\tilde{z} = z\,A/(2 \overline{\vform})$ are dimensionless time and space coordinates, where $\overline{\vform}$ is the time average value of $\vform(t)$. In these dimensionless coordinates, the solution is independent of $A$ and $\overline{\vform}$. (a) Formation only, occurring at a deposition front moving with variable speed; (b) Sequence of formation followed by resorption with net positive balance, resulting in discontinuities (``patches'').}\label{fig:ocy}
\end{figure*}
\section{Application to bone tissue}
Bone is a dynamic tissue that sustains lifelong changes in its microstructure and in its material properties~\cite{martin-burr-sharkey}. At the cellular scale, bone is composed of (i) bone matrix, infiltrated with minerals and with the osteocyte network; and (ii) vascular pores, containing soft tissues and cells. Changes in bone microstructure occur by dissolution of old bone matrix by bone-resorbing cells (osteoclasts) and deposition of new bone matrix by bone-forming cells (osteoblasts)~\cite{molec-biol-cell,martin-burr-sharkey,parfitt-in-recker}. Changes in material properties of newly deposited bone occur by matrix maturation such as collagen fiber re-arrangement, mineralisation, accumulation of micro-cracks, and maturation of osteocytes~\cite{martin-burr-sharkey,parfitt-in-recker}.

Bone remodelling turns over bone tissue slowly, at rates of 5--30\%/year. This allows bone matrix to undergo significant changes in material properties before being renewed. As a result, the state of bone is ``patchy'': it contains many internal boundaries separating tissue regions of different ages, which reflect the history of their formation and resorption processes. These different tissue regions are called bone structural units or osteons~\cite{martin-burr-sharkey,parfitt-in-recker,parfitt-1994}.

During bone modelling and remodelling, the bone surface $S(t)$ between bone matrix and vascular pores evolves by the action of osteoblasts and osteoclasts. The normal velocity of $S(t)$ is given by 
\begin{align}\label{form-res-velocity}
    \vform(\b r, t) = \kform\ \rho_\ob, \qquad \vres(\b r, t) = \kres\ \rho_\oc,
\end{align}
where $\rho_\ob(\b r, t)$ and $\rho_\oc(\b r, t)$ are the surface density of osteoblasts and osteoclasts (number per unit surface), $\kform(\b r, t)$ is the secretory rate (volume formed per \pbuenzli{osteoblast} per unit time), and $\kres(\b r, t)$ is the resorption rate (volume dissolved per \pbuenzli{osteoclast} per unit time)~\cite{buenzli2015}.

\subsubsection*{Osteocyte density}
Osteocytes are tissue-embedded cells believed to sense and transduce mechanical strains of bone matrix to osteoblasts and osteoclasts. Osteocytes \pbuenzli{reside} in small cavities and channels within bone matrix, making up a porosity of about 1--2\%~\cite{buenzli-sims}. No modelling or remodelling is initiated at these micropore surfaces. Osteocytes are generated with new bone matrix during formation. They can be viewed as a \pbuenzli{bone material property}, generated initially with density $\ocy_\form$. The spatio-temporal evolution of osteocyte density $\ocy$ (N.Ot/BV in bone histomorphometric standards~\cite{dempster-etal}) is governed by:
\begin{align}\label{ocy}
    \pd{}{t}\ocy(\b r, t) = \ocy_\form\ \vform\, \deltaup_{S(t)} - \ocy\ \vres\, \deltaup_{S(t)} - A\ \ocy.
\end{align}
The last term accounts for apoptosis (cell death) occurring with rate $A(\b r, t)$. In Ref.~\cite{buenzli2015}, a similar evolution equation for osteocyte density was proposed, but no resorption was accounted for. The first term was modelled as $\dburial\ \rho_\ob\ \deltaup_{S(t)}$ to represent the fact that osteocytes are osteoblasts that become buried during bone formation, where $\dburial(\b r, t)$ is the burial rate, \ie, the probability per unit time for an osteoblast to become \pbuenzli{trapped} in bone as an osteocyte. By identification with the term $\ocy_\form\, \vform\,\deltaup_{S(t)}$ in Eq.~\eqref{ocy}, one immediately finds that the density of osteocyte generated at the moving deposition front is given by
\begin{align}\label{ocygen}
    \ocy_\form(\b r, t) = \frac{\dburial\ \rho_\ob}{\vform} = \frac{\dburial}{\kform},
\end{align}
as obtained in~\cite{buenzli2015}. Equation~\eqref{ocygen} holds generally for the density of any inclusion deposited by osteoblasts in bone matrix at rate $\dburial$, and by extension, for any inclusion in tissue or material synthesised at an interface. This density does not explicitly depend on surface curvature and osteoblast density. This is particularly relevant for \pbuenzli{biological tissues and} biomaterials, since consistent inclusion densities can be generated in complex geometries and nonconstant populations of tissue-synthesising cells simply by maintaining the cell-specific properties $\dburial$ and $\kform$ constant.

Equation~\eqref{ocy} was solved numerically in one spatial dimension \pbuenzli{($z$)} when $A$ and $\ocy_\form$ are constant, and $v=v(t)$ oscillates between two values (Figure~\ref{fig:ocy}). In Figure~\ref{fig:ocy}a, $v(t)$ is always positive: there is no resorption. The solution surface in $(z,t)$ space is swept by a family of decreasing exponentials in time starting with value $\ocy_\form$ at the moving deposition front. The oscillation in front velocity generates nearby tissue regions (along $z$) that differ steeply, but continuously, in osteocyte density. Appendix~\ref{appx:numerical} contains details on the numerical  scheme and a comparison with the analytic solution
\begin{align}\label{ocy-theor}
    \ocy(\b r, t) = \mathbbm{1}_{\bv(t)}(\b r)\ \ocy_\form(\b r, t)\,
\exp\left\{- \int_{t^\ast(\b r)}^t\hspace{-1em}\der t\ A(\b r, t)\right\},
\end{align}
where $t^\ast(\b r)$ is the arrival time at $\b r$, $\bv(t)$ is the spatial region occupied by bone at time $t$, and $\mathbbm{1}_{\bv(t)}$ is the indicator function of $\bv(t)$. \pbuenzli{This solution was derived in Ref.~\cite{buenzli2015} like Eqs~\eqref{m-gov-eq}--\eqref{m-sol} for bone mineral density below.}

In Figure~\ref{fig:ocy}b, $v(t)$ oscillates between a positive and a negative value: there is an alternation of bone tissue formation and bone tissue resorption. Resorption introduces sharp discontinuities in osteocyte density in adjacent regions, resulting in a bone matrix composed of distinct tissue layers (``patches''). These patches are due to the fact that tissue lying under resorbing surfaces keep maturing. When resorption stops and new tissue forms, there is an age gap between the underlying tissue and new tissue.

The analytic solution~\eqref{ocy-theor} holds within each patch region, which may shrink during resorption. The solution in the whole space can be constructed piecewise. However, this requires book-keeping of the time and locations at which there is reversal between resorption and formation to identify patches. Such book-keeping is tedious and impractical in higher dimensions as tissue formation events may be generated at different times and locations of the surface. The governing equation~\eqref{ocy} can represent these patches without explicitly needing the information of resorption--formation reversals. It can also handle more elaborate situations, such as nonlinearities and complex couplings.

\subsubsection*{Bone mineral density}
New bone is formed initially as an unmineralised collagen matrix. This unmineralised matrix matures and gradually incorporates minerals to become hard bone tissue. Mineralisation first increases rapidly due to the deposition of mineral pellets by cells during formation. It then continues to increase over much larger time scales by crystal growth~\cite{parfitt-in-recker}. Mineral density is an important bone material property. It is measured clinically as an indicator of skeletal integrity, for example in osteoporosis~\cite{briggs-perilli-etal-2010}. 
Assuming that new bone tissue is infiltrated with an initial density of mineral pellets $m_\form(\b r, t)$, the spatio-temporal evolution \pbuenzli{of} bone mineral density is governed by:
\begin{align}\label{m-gov-eq}
    \pd{}{t}m(\b r, t) = m_\form\,\vform\,\deltaup_{S(t)} - m\,\vres\,\deltaup_{S(t)} + \Fminer(m).
\end{align}
The mineralisation law $\Fminer$ determines the evolution of mineral density after the initial pellet deposition. Without resorption, $m(\b r, t)$ is solution of the initial value problem
\begin{align}
    &\pd{}{t}m(\b r, t) = \Fminer(m),\quad \forall t>t^\ast(\b r),\label{mineralisation}
    \\&m\big(\b r, t^\ast(\b r)\big) = m_\form.\label{mform}
\end{align}
The initial value~\eqref{mform} expresses the jump property $\eqref{jump}$ at time $t=t^\ast(\b r)$ due to the surface balance term $m_\form\,\vform\,\deltaup_{S(t)}$: at $t=t^\ast(\b r)$, $m$ jumps from $0$ to $m_\form$. After the initial mineral deposition, we assume that bone mineral density increase until it reaches a maximum mineral density $\mmax$. We model this mineralisation process by exponential saturation:
\begin{align}\label{miner-law}
    \Fminer(m) = 
    \begin{cases}
            0, & \quad \text{if $m=0$}, 
            \\\kminer (\mmax - m), & \quad \text{if $m>0$}.
    \end{cases}
\end{align}
If $\mmax(\b r)$ is independent of time in Eq.~\eqref{miner-law}, the solution to~\eqref{mineralisation}--\eqref{miner-law} is:
\begin{align}\label{m-sol}
    m(\b r, t) = \mmax - &\left(\mmax - m_\form\right)
\exp\left\{-\int_{t^\ast(\b r)}^t\hspace{-1em}\der t'\ \kminer(\b r, t')\right\},
\end{align}
where $m_\form$ is evaluated at $m_\form\left(\b r, t^\ast(\b r)\right)$. In reality, $\mmax$ is likely to be a function of time. It is believed to be regulated by osteocytes and their \pbuenzli{dendritic} processes~\cite{parfitt-in-recker,kerschnitzki-2013}.

\begin{figure}[t] \centering\includegraphics[width=\figurewidth]{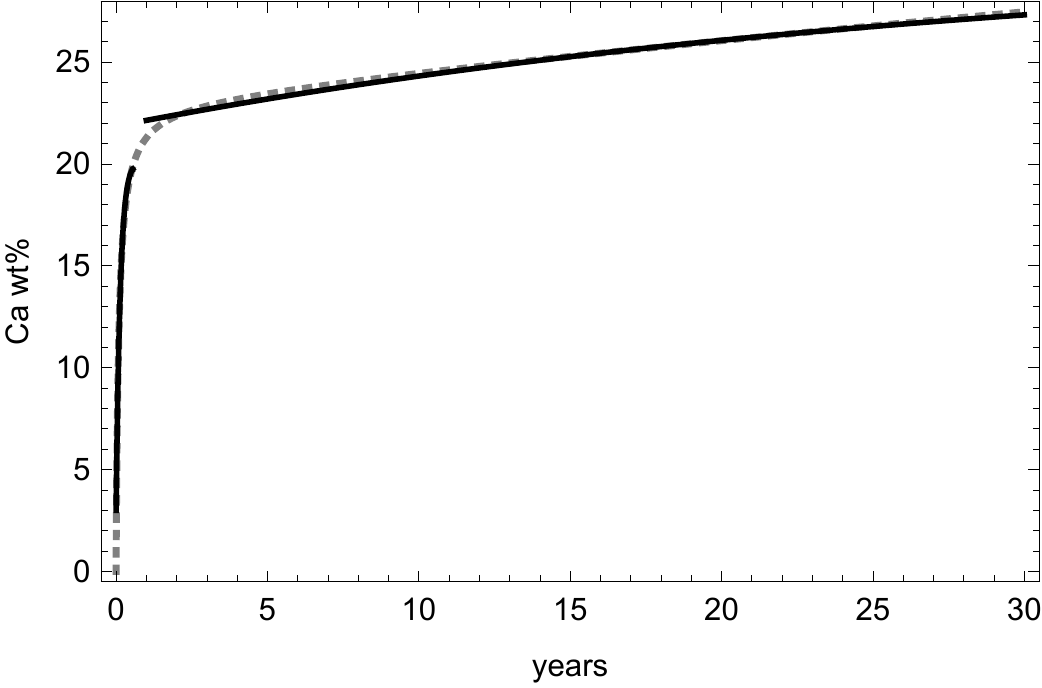}
\begin{picture}(0,0)
\put(-40,45){\includegraphics[width=0.6\figurewidth]{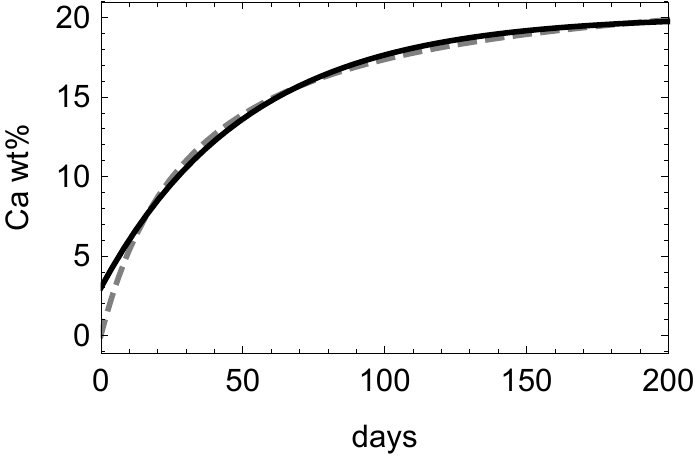}}
\end{picture}
\caption{Experimental mineralisation kinetics from Ref.~\cite{ruffoni-etal-2007} (dashed) together with two exponential fits of the form $\mmax \!- (\mmax\!-\! m_\form)\,\e^{-\kminer t}$ (solid lines). The short-time fit has $m_\form=3$ Ca wt\%, $\mmax = 20.1$ Ca wt\%, $\kminer = 0.0194/\da$ (also shown in the inset). The long-time fit has $m_\form=21.85$ Ca wt\%, $\mmax = 30.4$ Ca wt\%, $\kminer = \num{9.3e-5}/\da$.}
\label{fig:miner-fits}
\end{figure}

\begin{figure*}[t]
\centering\includegraphics[width=\textwidth]{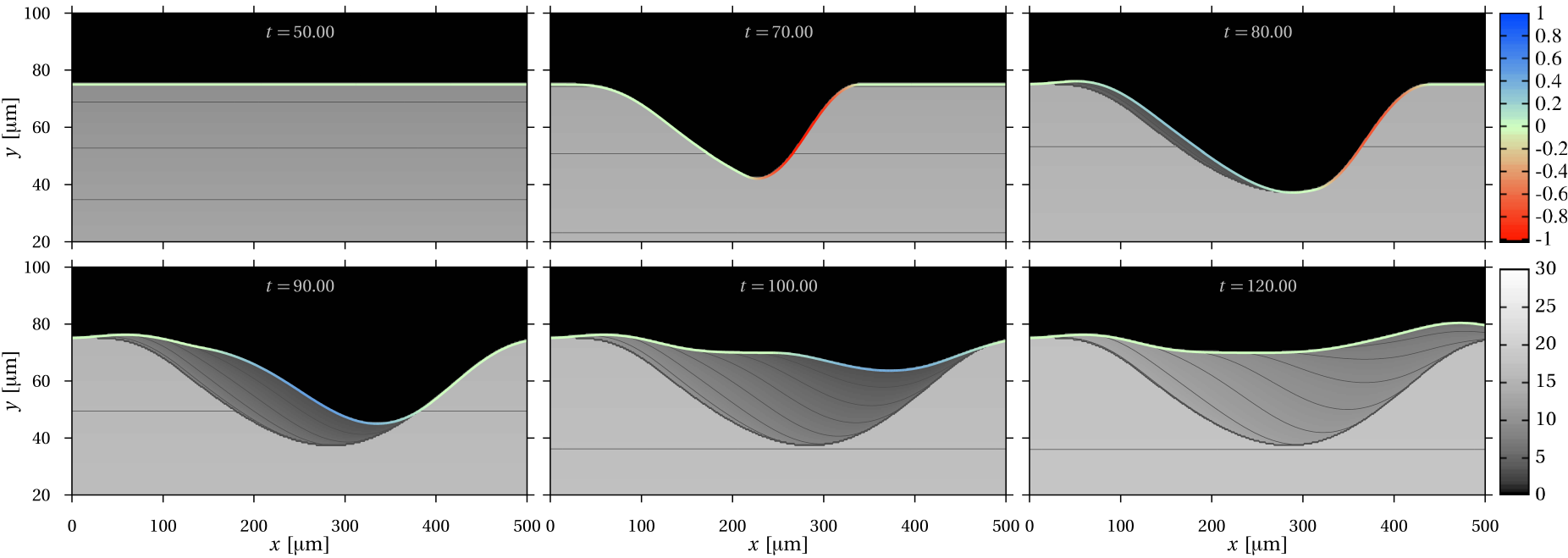}
\caption{Time snapshots of the evolution of bone mineral density in a portion of bone undergoing remodelling. The grey scale of the bone matrix represents mineral density in units of Ca wt\%~\cite{ruffoni-etal-2007} with contour lines shown every 1 Ca wt\%. The color scale of the interface is the normal velocity, normalised by the maximum absolute value in this simulation
. At $t=50$\,days, remodelling is initiated with osteoclasts (red) starting to resorb bone matrix until $t=90$\,days. At $t=70$\,days, osteoblasts (blue) are activated towards the rear and start refilling the resorbed cavity until $t=120$\,days, at which point the interface is still and remodelling has completed. Because bone tissue resorption has removed a portion of bone adjacent to mineralising tissue, newly formed bone contrasts by its mineral content with surrounding tissue. The end state of the bone matrix is made up of two distinct patches, within which mineral density is continuous.}\label{fig:miner1}
\end{figure*}
Experimental determinations of the increase in mineral density with time in newly deposited bone tissue exhibit two time scales (Figure~\ref{fig:miner-fits})~\cite{ruffoni-etal-2007}. While explicit fitting functions for $t\mapsto m(\b r, t)$ have been proposed with great accuracy to experimental data~\cite{ruffoni-etal-2007}, these fitting functions do not satisfy a simple mineralisation kinetics law of the type~\eqref{mineralisation}.
We assume instead that mineralisation is described by the exponential saturation law~\eqref{miner-law} with distinct characteristic times at these two time scales. The constants $m_\form$, $\mmax$, and $\kminer$ in Eq.~\eqref{miner-law} are adjusted to fit the experimental mineralisation kinetics of Ref.~\cite{ruffoni-etal-2007} either at short times ($10$--$200$ days; $m_\form=3$\,Ca\,wt\%, $\mmax=20.1$\,Ca\,wt\%, $\kminer=0.0194$/day) or at large times (1--30 years; $m_\form=21.85$\,Ca\,wt\%, $\mmax=30.4$\,Ca\,wt\%, $\kminer=\num{9.3e-5}$/\da), see Figure~\ref{fig:miner-fits}.

Figure~\ref{fig:miner1} shows time snapshots of a simulated bone remodelling event in two-dimensional space operated by a succession of bone-resorbing cells and bone-forming cells. The cell populations were assigned so as to emulate a transient travelling wave of bone renewal representing a basic multicellular unit ($\bmu$)~\cite{parfitt-1994}. Bone mineral density was evolved using the short-time mineralisation parameters. Bone-resorbing cells first create a cavity in a mineralising bone tissue substrate. Bone-forming cells then deposit new tissue. The new tissue contrasts with the older substrate by its lower mineral content. After remodelling has completed, the tissue is clearly made up of two distinct patches. Within each patch, the mineral density keeps increasing and is continuous, but it is discontinuous at the line corresponding to the deepest location reached by resorption. In bone, this line of reversal between resorption and formation is called the cement line. The patch of newly formed bone is called a secondary osteon, or bone structural unit~\cite{parfitt-in-recker}. 

The overall bone balance after the remodelling event in Figure~\ref{fig:miner1} is approximately zero. However, the interface has changed. Small changes in the interface are likely to occur in bone remodelling even without bone loss or gain. Indeed, bone remodelling is regulated by several processes of biochemical, geometrical, and mechanical nature, which affect the generation and coupling of bone-resorbing and bone-forming cells. These regulatory processes were not modelled here, see Refs~\cite{ryser-etal1,ryser-etal2,buenzli-etal-moving-bmu,buenzli-etal-bmu-refilling} for more biologically accurate mathematical models of cell population dynamics in $\bmu$s.

\begin{figure}[t]
\centering\includegraphics[width=0.9\figurewidth]{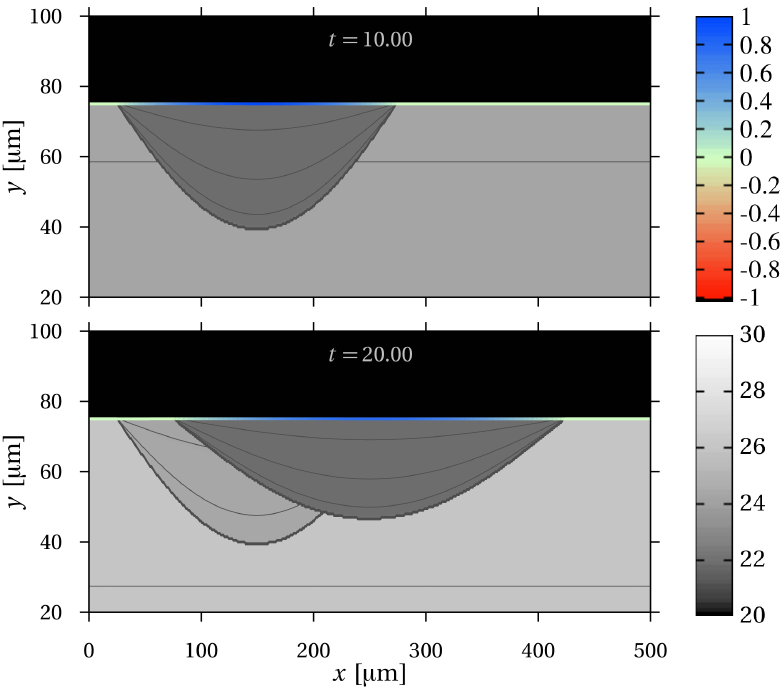}
\caption{Time snapshots of bone mineral density in a portion of bone after a first and a second remodelling event, occurring at $t=10$\,years and $t=20$\,years. Contour lines are shown every 0.02 Ca wt\%. Note the different grey scale compared to Figure~\ref{fig:miner1}. Part of the record of the first remodelling event has been overwritten by the second remodelling event.}\label{fig:miner2}
\end{figure}
Figure~\ref{fig:miner2} shows a portion of bone that underwent two bone remodelling events in twenty years, roughly  corresponding to a turnover rate of 10\%/year~\cite{martin-burr-sharkey,parfitt-in-recker}. 
\pbuenzli{The cell populations were assigned so as to emulate remodelling events without net bone gain or loss (Appendix~\ref{appx:numerical}).} Bone mineral density was evolved using the long-time mineralisation parameters. After the second remodelling event, the tissue is made up of three distinct patches: the old bone substrate, bone renewed by the first remodelling event, and bone renewed by the second remodelling event. Part of the bone renewed by the first remodelling event was removed and replaced by newer bone during the second remodelling event. This kind of variegated state of bone matrix is typical, as observed by microradiographs~\cite{cohen-harris,tappen,robling-stout}, quantitative back-scattering electron microscopy~\cite{ruffoni-etal-2007,lukas-etal-2011}, and micro-computed tomography~\cite{arhatari,sansalone}. Quantities recorded in bone during formation are gradually overwritten with newer content. This constitutes a loss of information: osteocyte density for example records the ratio of burial rate to secretory rate that is current at the time of formation, see Eq.~\eqref{ocygen}~\cite{buenzli2015}. On the other hand, bone tissue patches provide other information such as the age or turnover rate of the tissue.

As in the one-dimensional simulation, the governing equation~\eqref{m-gov-eq} can represent tissue patches without needing the information of the time and locations of resorption--formation reversals. Book-keeping patch location in space and time is particularly complicated in situations such as Figure~\ref{fig:miner2} where previous reversal surfaces are partially erased.

\subsection{Bone tissue spatial averages}
Equations~\eqref{avg2}--\eqref{avg3} are valid in general. Here we specialise them to bone tissue using Eqs~\eqref{form-res-velocity} and further assume that the secretory rate $\kform$ and dissolution rate $\kres$ are constant. We also denote $V$ by \tv\ (tissue volume) and $\Omega(t)$ by \bv\ (bone volume) to follow bone histomorphometric conventions~\cite{dempster-etal}. \pbuenzli{Under these assumptions:}
\begin{align}
&\pbuenzli{\frac{\sform}{V}\avgg{\vform}{\sform} =\frac{\sform}{V} \kform\avgg{\rho_\ob}{\sform} = \kform\,\frac{1}{V}\!\!\int_\sform\hspace{-0.5em} \der\sigma\,\rho_\ob = \kform \frac{\text{N}.\ob}{V} =\kform \avgg{\ob}{\tv}}\notag
\\& \pbuenzli{\frac{\sres}{V}\avgg{\vres}{\sres} =\frac{\sres}{V} \kres\avgg{\rho_\oc}{\sres} = \kres\,\frac{1}{V}\!\!\int_\sres\hspace{-0.5em} \der\sigma\,\rho_\oc = \kres \frac{\text{N}.\oc}{V} =\kres \avgg{\oc}{\tv},}\label{bone-averages}
\end{align}
\pbuenzli{where $\text{N}.\ob$ and $\text{N}.\oc$ are the number of osteoblasts and osteoclasts in $\tv$. Using Eqs~\eqref{bone-averages} in Eq.~\eqref{f-gov-eq}}, bone volume fraction evolves as:
\begin{align}\label{fbm-gov-eq}
    \td{f(t)}{t} = \kform \avgg{\ob}{\tv} - \kres \avgg{\oc}{\tv}.
\end{align}
Equation~\eqref{fbm-gov-eq} provides a microscopic justification of the equation
$\td{f(t)}{t} = \kform\ \ob - \kres\ \oc$ used in the literature, where $\ob$ and $\oc$ are average cell densities in a representative elementary volume~\cite{martin1972,pivonka-2008,buenzli-etal-endocortical-2013,lerebours-bmmb}. 

\subsubsection*{Mean-field approximation of bone mineral density}\label{sec:mean-field}
Current conventional microCT scanners have millimetric to submillimetric resolution. They effectively measure local spatial averages of bone mineral densities. If soft tissues are included in the average, measurements refer to `bone mineral density' ($\bmd$). If soft tissues are excluded, measurements refer to `tissue mineral density' ($\tmd$) ~\cite{bouxsein-mueller}. Thus:
\begin{align}
    \avgg{m}{\tv} = \bmd, \quad \avgg{m}{\bv} = \tmd.
\end{align}

The evolution of $\avgg{m}{\tv}$ and $\avgg{m}{\bv}$ by Eqs~\eqref{avg2} and \eqref{avg3} depends on the patchy state of bone, and so on remodelling history. However, if bone mineral density is not too inhomogeneous in $\bv$, we can close Equations~\eqref{avg2} and \eqref{avg3} by making the \emph{mean-field approximation}:
\begin{align}\label{mean-field}
    m \approx \avgg{m}{\bv}.
\end{align}
With the mineralisation model used in Figures~\ref{fig:miner-fits}--\ref{fig:miner2}\pbuenzli{, which assumes $k_f$, $k_r$, $k_m$, and $m_\form$ constant, we have:}
\begin{align*}
    &\pbuenzli{\frac{\sform}{V}\avgg{\vform m_\form}{\sform} = m_\form \frac{\sform}{V}\avgg{\vform}{\sform} = m_\form \kform\avgg{\ob}{\tv},}
    \\&\pbuenzli{\frac{\sres}{V}\avgg{\vres m}{\sres} \approx \frac{\sres}{V}\avgg{\vres\avgg{m}{\bv}}{\sres} = \avgg{m}{\bv} \frac{\sres}{V}\avgg{\vres}{\sres} = \avgg{m}{\bv} \kres\avgg{\oc}{\tv},}
\end{align*}
\pbuenzli{where the last equalities in each line used Eqs~\eqref{bone-averages} and the first equality in the second line used the mean-field approximation~\eqref{mean-field}. Eq.~\eqref{avg2} thus becomes:}
\begin{align}
    \td{\avgg{m}{\tv}}{t} \approx & m_\form \kform \avgg{\ob}{\tv} - \avgg{m}{\bv}\kres \avgg{\oc}{\tv} + \Big\langle\Fminer\Big(\avgg{m}{\bv}\Big)\Big\rangle_{\!\!\tv} \notag
    \\= & m_\form \kform \avgg{\ob}{\tv} - \tfrac{1}{f} \avgg{m}{\tv}\kres \avgg{\oc}{\tv} + \kminer\left(f \mmax - \avgg{m}{\tv}\right),\label{bmd}
\end{align}
and Eq.~\eqref{avg3} becomes:
\begin{align}
    \td{\avgg{m}{\bv}}{t} \approx & \tfrac{1}{f}\left(m_\form - \avgg{m}{\bv}\right) \kform\avgg{\ob}{\bv} \notag
\\ & - \tfrac{1}{f}\left(\avgg{\avgg{m}{\bv}}{\sres}-\avgg{m}{\bv}\right)
\kres\avgg{\oc}{\bv}
+ \Big\langle\Fminer\Big(\avgg{m}{\bv}\Big)\Big\rangle_{\!\!\bv}\notag
\\ = &\tfrac{1}{f}\left(m_\form - \avgg{m}{\bv}\right)\kform \avgg{\ob}{\bv} + \kminer\left(\mmax - \avgg{m}{\bv}\right).\label{tmd}
\end{align}
For given average densities of osteoblasts and osteoclasts, $f$ is given by Eq.~\eqref{fbm-gov-eq}, and Eqs~\eqref{bmd}, \eqref{tmd} are now self-consistent.

The time evolution of $\avgg{m}{\bv}$ and $\avgg{m}{\tv}$ found by explicitly averaging the microscopic model~\eqref{m-gov-eq}, \eqref{miner-law} exhibits specific model elements in the different terms of Eqs~\eqref{bmd} and~\eqref{tmd}. These elements could easily be missed when heuristically formulating a temporal model directly:
\begin{enumerate}
    \item[(i)] The factor $1/f$ in Eq.~\eqref{bmd} is due to Eq.~\eqref{avgV-avgT}; 
    \item[(ii)] The factor $f$ multiplying $\mmax$ in Eq.~\eqref{bmd} is due to the fact that $\Fminer(m)$ is not linear in $m$; it is dicontinuous at $m=0$. In fact, $\Fminer$ is such that $\mmax=0$ out of $\bv$, so that $\avgg{\mmax}{\tv} =  f\mmax$.
    \item[(iii)] The evolution of $\avgg{m}{\bv}$ in Eq.~\eqref{tmd} is independent of resorption. The dependence on formation corresponds to the relaxation of $\avgg{m}{\bv}$ towards the value deposited $m_\form$. The relaxation rate is proportional to the bone formation rate $\kform \avgg{\ob}{\bv}$ and to $1/f$. The lower the bone volume fraction $f$, the quicker it is to replace the current average $\avgg{m}{\bv}$ with new values $m_\form$.
\end{enumerate}

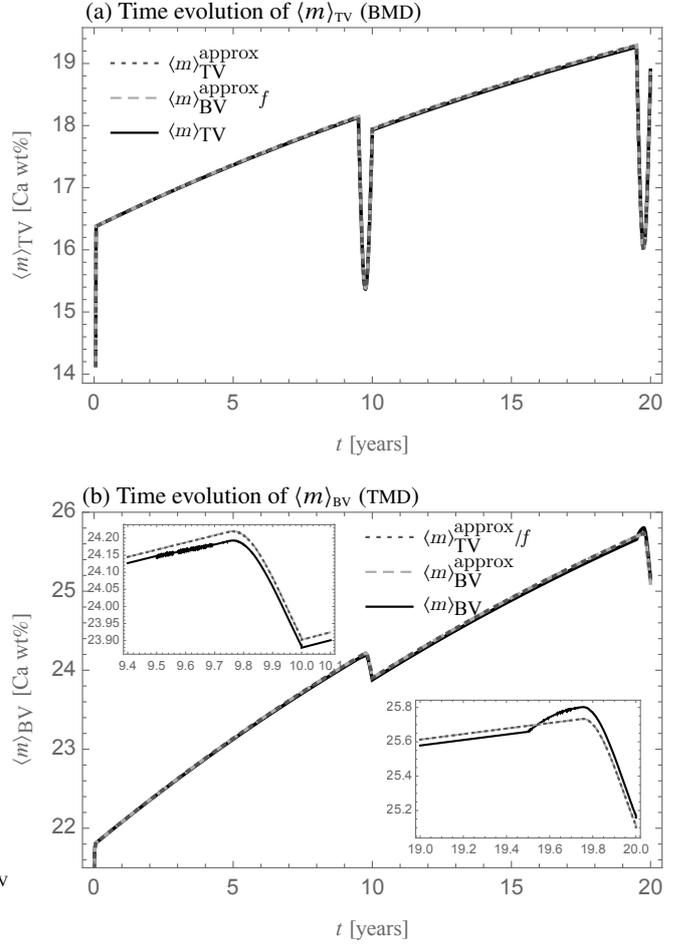
\begin{figure}[t]\vspace*{2ex} \centering
\includegraphics[width=\figurewidth]{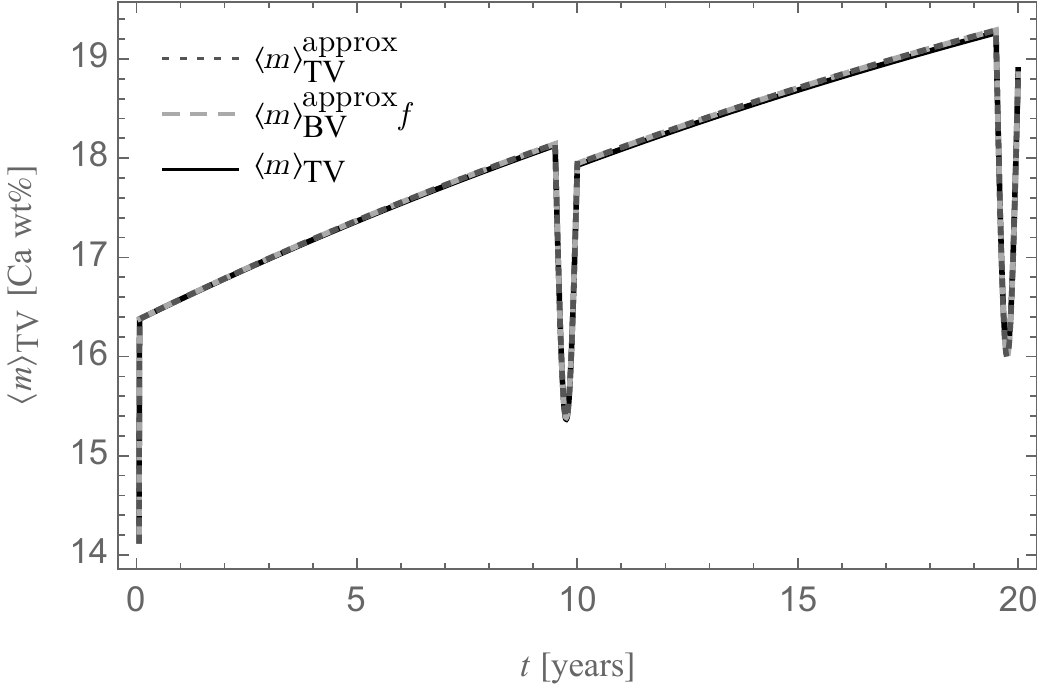}
\begin{picture}(0,0)
\put(-94,180){\small (a) Time evolution of $\avg{m}_\tv$ ($\bmd$)}
\end{picture}
\includegraphics[width=\figurewidth]{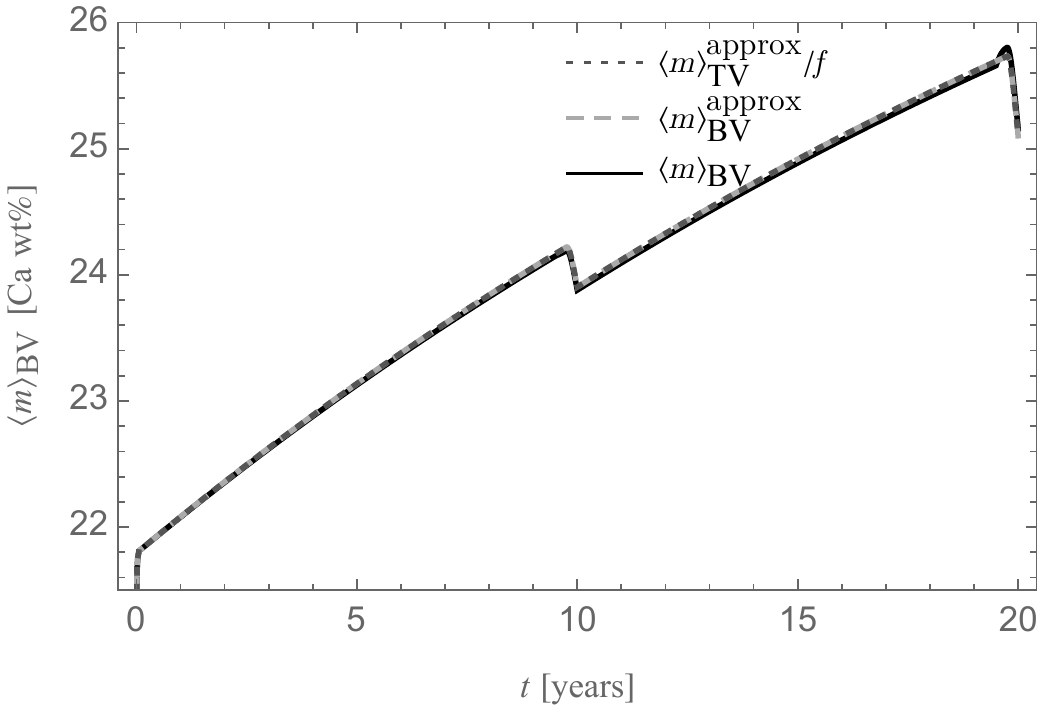}
\begin{picture}(0,0)
\put(-94,180){\small (b) Time evolution of $\avg{m}_\bv$ ($\tmd$)}
\put(-92,115){\includegraphics[width=0.39\figurewidth]{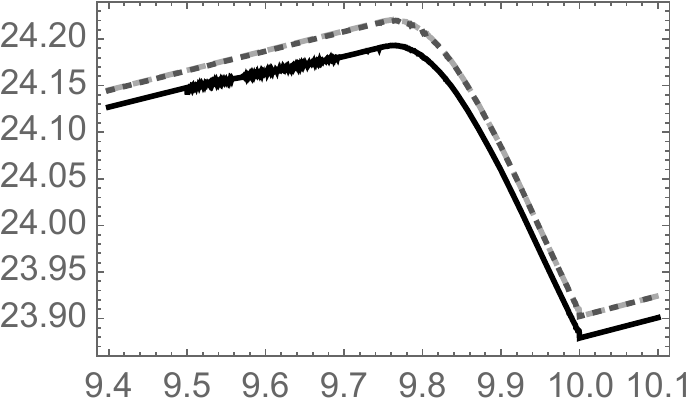}}
\put(20,45){\includegraphics[width=0.4\figurewidth]{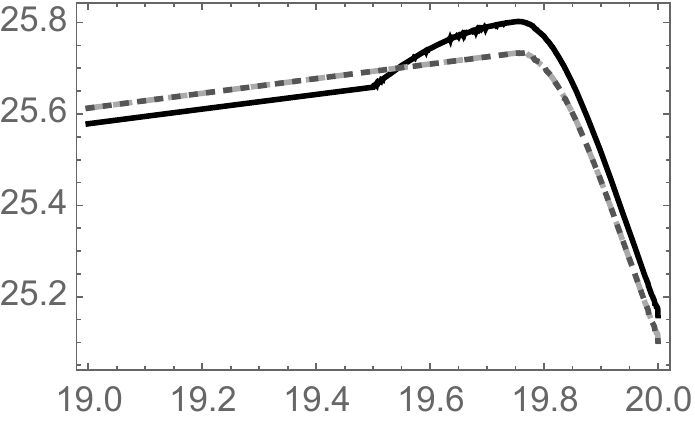}}
\end{picture}
\vspace*{-5mm}
\caption{Comparison between different evaluations of bone mineral density averages during the two remodelling events of Figure~\ref{fig:miner1}. Solid line: based on the spatio-temporal solution; interrupted lines: based on the mean-field approximation, Eqs~\eqref{bmd} and Eq.~\eqref{tmd}. (a) Evolution of $\avgg{m}{\tv}$; (b) Evolution of $\avgg{m}{\bv}$. The insets close up on the behaviour around the remodelling times.}
\label{fig:bmd-averages}
\end{figure}
Figure~\ref{fig:bmd-averages} shows the time evolution of the spatial averages $\avgg{m}{\tv}$ ($\bmd$) and $\avgg{m}{\bv}$ ($\tmd$) during the two remodelling events simulated in Figure~\ref{fig:miner2}. The remodelling events occur at $t=9.5$--$10\,\text{years}$ and $t=19.5$--$20\,\text{years}$. The solid line is based on numerically integrating the spatio-temporal numerical solution shown in Figure~\ref{fig:miner2}. The interrupted lines are based on the solutions of the mean-field ODEs~\eqref{bmd} and~\eqref{tmd}\pbuenzli{, in which the average cell densities $\avgg{\ob}{\tv}$ and $\avgg{\oc}{\tv}$ are integrated from the given spatio-temporal expressions for cell densities used in Figure~\ref{fig:miner2} (see Appendix~\ref{appx:numerical})}.

Except during the remodelling events, $\avgg{m}{\tv}$ and $\avgg{m}{\bv}$ increase due to the mineralisation law~\eqref{miner-law}. The large dips in $\avgg{m}{\tv}$ are due to the changes in $f$ during the remodelling events. Both $\avgg{m}{\tv}$ and $\avgg{m}{\bv}$ have decreased values at the end of each remodelling event compared to the value prior to the remodelling event. This is due to the presence of new, lower-mineralised bone after remodelling. The numerical solutions of the mean-field equations~\eqref{bmd} and~\eqref{tmd} (scaled by either $f$ or $1/f$ according to Eq.~\eqref{avgV-avgT}) are undistinguishable. The mean-field solutions differ from the averaged spatio-temporal solution by 0.01\% at the first remodelling event, and by 0.04\% at the start of the second remodelling event (see insets). These differences are attributed to the different numerical integrations required by the solutions. However, if model elements listed in points (i)--(iii) above are missed, the mean-field solutions can differ dramatically (not shown).

Just before the first remodelling event, bone mineral density is approximately homogenous. When this bone is remodelled, there is no qualitative difference between the averaged spatio-temporal solution and the mean-field approximations (top-left inset). However, just before the second remodelling event, bone mineral density is distributed heterogenously across two patches (Figure~\ref{fig:miner2}a). When this bone is remodelled, the average mineral density at the surface is significantly lower than the bone volume average. The removal of bone near the surface during resorption thus accelerates the increase in $\avgg{m}{\bv}$ in a first stage (Figure~\ref{fig:bmd-averages}b, solid line in bottom-right inset), before $\avgg{m}{\bv}$ decreases due to new, lower-mineralised bone being deposited during formation. This initial accelerated increase in $\avgg{m}{\bv}$ is missed by the mean-field approximations (interrupted lines).

\pbuenzli{The behaviours of $\bmd$ and $\tmd$ around $t=10\,\yr$ and $t=20\,\yr$ in Figure~\ref{fig:bmd-averages} represent typical signatures of isolated remodelling events that could in principle be seen by \emph{in-vivo} microCT of resolution $\tv$. These time signatures may therefore provide a way to detect remodelling events that occur at lower, microscopic spatial resolutions not seen in the scans. Current \emph{in-vivo} technologies remain limited in the number of timepoints and the transient behaviours during the remodelling events may be missed. However, sawtooth-like changes in mineral density may still be detected. This would require a resolution-accurate co-registration of scans taken at different timepoints. Note that concurring remodelling events within a voxel $\tv$ could smear out individual remodelling signatures.}

\section{Conclusions}
This paper shows that the evolution of tissue geometry and tissue material properties under modelling and remodelling processes can be captured by a single, general mathematical framework. Tissue heterogeneities due to different tissue ages and different biological contexts at creation are represented in this framework by functions of space and time with discontinuities at internal or external boundaries. The equations governing the evolution of these functions are singular differential equations, in the sense of distribution theory. The surface distribution $\deltaup_S$ and associated jump property~\eqref{jump} enable a modular approach to formulating governing equations of complex tissues and biomaterials. They extend conventional balance equations with nonconservative processes localised at moving boundaries. This enables `continuum model' notations to be employed despite the occurrence of discontinuities at surfaces, much like the Dirac distribution enables continuum notations to be employed in discrete systems~\cite{evans-morriss}. 

Internal tissue boundaries that separate regions generated at different times are created at reversals \pbuenzli{between resorption and formation}. These boundaries arise naturally from the governing equation~\eqref{gov-eq}. In contrast, analytical solutions require to book-keep the time and location of \pbuenzli{these} reversals to construct the solution piecewise from continuous patches.

A distinction is sometimes made in biology between tissue modelling and tissue remodelling. From the point of view of following the evolution of tissue properties, these processes do not need to be distinguished, so long as the effect of removal and formation on tissue properties are identical in both situations. This is the case of bone tissues, for which remodelling can be seen as a coordinated sequence of small resorptive and formative modelling processes~\cite{martin-burr-sharkey}. The evolution of bone tissue during modelling or remodelling is thus described mathematically by the same set of equations, the difference being in the timing and location of the resorption and formation processes. Here, these were assumed given. In practice, this information may come from experimental data, or from further mathematical models of the populations of bone-resorbing and bone-forming cells. Generally, the governing equation~\eqref{gov-eq} needs to be supplemented with information on the specific processes involved in the formation and resorption kinetics of the tissue, which determine the normal velocities $\vform$ and $\vres$, and the value $\eta_\form$ of newly formed tissues.

Exact governing equations for the evolution of spatial averages of the tissue were obtained by integrating the spatio-temporal equation~\eqref{gov-eq}. These average equations are not self-consistent due to the heterogeneous nature of the tissue, but can be closed by a mean-field approximation such as Eq.~\eqref{mean-field}. The degree to which the mean-field approximation is well satisfied depends on the degree of inhomogeneity of the tissue. Caution should be exercised whenever a tissue property changes over time scales that are faster than typical remodelling rates, which results in patchy states such as in Figures~\ref{fig:miner1}, \ref{fig:miner2}. While the discrepancy due to the mean-field approximation in Figure~\ref{fig:bmd-averages}b is small, such discrepancies would accumulate with further remodelling events. \pbuenzli{Figure~\ref{fig:bmd-averages} is a prediction of the type of $\bmd$ or $\tmd$ signatures that could be detected by \emph{in-vivo} microCT scans when the bone undergoes remodelling at lower, unseen length scales.}

The spatial and temporal scales at which the formalism presented in this paper is valid depend on the adequacy of the continuum model to represent a particular application at these scales. The bone tissue examples presented here were considering boundaries to be the bone--vascular interface. At a lower scale, boundaries may represent the secretory areas of a cell's membrane. At a higher scale, boundaries may represent the overall shape of an organ. This formalism is applicable to many other systems in which a material is created and destroyed from its surfaces \pbuenzli{while} undergoing changes in the bulk. This includes tissues and biomaterials such as ECM remodelling, tooth development, the generation and biomineralisation of shells, \pbuenzli{bioscaffolds,} but also non-biological systems, such as sedimentation, 3D printing, etching, and chemical adsorption.

\subsection*{Acknowledgements}
I would like to thank Richard Weinkamer, Chlo\'e Lerebours, and Jake Taylor-King for stimulating discussions. I gratefully acknowledge the Australian Research Council for Discovery Early Career Researcher Fellowship (project number DE130101191).

\begin{appendices}
\section{Properties of the surface distribution}\label{appx:surface-distrib-properties}
This appendix presents a few properties of the surface distribution $\deltaup_S$  defined by Eq.~\eqref{def-surface-distrib}~[\citealp{bedeaux-albano-mazur-1976,ronis-bedeaux-oppenheim-1978,albano-bedeaux-vlieger-1979}; \citealp[Sec.~8.4]{jones-distrib}]. Intuitively, the surface distribution is similar to the Dirac distribution except that it is formally infinite on a $N-1$ manifold embedded in $\mathbb{R}^N$. Such a manifold is usually called a `hypersurface'. We will refer to it as a `surface' for simplicity. It corresponds to a curve when $N=2$ and a point when $N=1$. Integrating the surface distribution $\deltaup_S$ over $N$-dimensional space with a test function only retains the function's values on the $N-1$ dimensional surface, and integrates these values with respect to the measure defining $N-1$ dimensional area~\cite{makarov-podkorytov}. It is important to contrast the surface distribution with the Dirac distribution, which in all dimensions returns the value of a test function at a single point. We refrain from using the terminology `Dirac' to refer to the surface distribution $\deltaup_S$ to avoid potential confusion. In Section~\ref{appx:curvilinear-partition}, we first demonstrate the local curvilinear partition of space, Eq.~\eqref{curvilinear-partition}. This relation means that in effect, integrating over space with $\deltaup_S$ removes spatial components normal to the surface $S$, see Eq.~\eqref{surface-distrib-arc-length}. In Section~\ref{appx:surface-distrib-representations} we mention several different representations of the surface distribution found in the literature. Finally, we mention applications to the evolution equation of an evolving domain's indicator function (Sec.~\ref{appx:indicator-gov-eq}) and to the representation of volumetric density of surface-bound quantities (Sec.~\ref{appx:volumetric-surface-density}).

\subsection{Jump property}\label{appx:jump-property}
We first demonstrate the jump property enunciated in Eq.~\eqref{jump}.

\noindent\textit{Proof.} In one spatial dimension, the interface is a point of coordinate $S(t)$. Eq.~\eqref{jump} is obtained by integrating Eq.~\eqref{jump-eq} over $t\in[t^\ast_-, t^\ast_+]$ and by using $\deltaup_{S(t)}(x) = \deltaup\big(x-S(t)\big) = \frac{1}{|v|}\deltaup(t - t^\ast)$, where $|v| = |S'(t^\ast)|$. To prove the jump property in higher dimensions, we first replace the running time variable $t$ in the right hand side of Eq.~\eqref{jump-eq} by $t^\ast$: only the values of $\Delta\eta$ and $v$ at the arrival time $t^\ast$ contribute to the change in $\eta$ at $\b r$. Indeed, for any function $\varphi(\b r, t)$:
\begin{align}
    &\int\hspace{-0.2em}\der \b r\ \deltaup_{S(t)}(\b r)\, \varphi(\b r, t) = 
    \int_{S(t)}\hspace{-1em}\der\sigma(\b r)\ \varphi(\b r, t) \notag
\\&= \int_{S(t)}\hspace{-1em}\der\sigma(\b r)\ \varphi\big(\b r, t^\ast(\b r)\big) 
= \int\hspace{-0.2em}\der\b r\ \deltaup_{S(t)}(\b r)\, \varphi\big(\b r, t^\ast(\b r)\big),\label{time-is-arrival-time}
\end{align}
where the second equality in Eq.~\eqref{time-is-arrival-time} uses the fact that any point $\b r\in S(t)$ has the arrival time $t^\ast(\b r)=t$. 

The pointwise notation in Eq.~\eqref{jump-eq} is elucidated in the theory of distributions by integrating over space with a smooth kernel function $\deltaup_n$ of unit integral, and of support tending to the single point $\{\b r\}$ as $n\to\infty$. The sequence $\deltaup_n$ is called a regular sequence converging to Dirac's delta distribution~\cite{jones-distrib}: $\lim_{n\to\infty}\deltaup_n (\b r'-\b r)= \deltaup(\b r' - \b r)$. The meaning of Eq.~\eqref{jump-eq} is thus
\begin{align}\label{jump-deriv1}
    \pd{}{t}\eta(\b r, t) = \lim_{n\to\infty}\int\hspace{-0.2em}\der \b r'\ \deltaup_n(\b r'-\b r)\, \Delta\eta(\b r', t^\ast)\,v(\b r', t^\ast)\, \deltaup_{S(t)}(\b r').
\end{align}
To calculate the jump in $\eta$ induced by the passage of $S(t)$ through $\b r$ at $t=t^\ast$, we integrate Eq.~\eqref{jump-deriv1} over $t\in[t^\ast_-, t^\ast_+]$ and use the definition~\eqref{def-surface-distrib}
\begin{align}\label{jump-deriv2}
    \eta&(\b r, t^\ast_+) - \eta(\b r, t^\ast_-)
\\&= \lim_{\epsilon\to 0}\lim_{n\to\infty}\int_{t^\ast_-}^{t^\ast_+}\hspace{-1em}\der t \!\!\int_{S(t)}\hspace{-1.2em}\der\sigma(\b r')\ \deltaup_n(\b r'- \b r)\, \Delta\eta(\b r', t^\ast)\, v(\b r', t^\ast)\notag.
\end{align}
The final step consists in partitioning space in a neighbourhood of $\b r$ by a set of parallel and perpendicular coordinates to the interface $S(t^\ast)$, such that
\begin{align}\label{curvilinear-partition}
    \der \b r' = |v|\der t\der\sigma.
\end{align}
This partioning is visually intuitive (see Figure~\ref{fig:eta-jump}). It is proved in Sec.~\ref{appx:curvilinear-partition}. With Eq.~\eqref{curvilinear-partition}, we finally obtain
\begin{align}
    \eta(\b r, t^\ast_+) - \eta(\b r, t^\ast_-) 
&= \lim_{\epsilon\to 0}\lim_{n\to\infty} 
\int_{V_{\epsilon}(\b r, t^\ast)}\hspace{-2em}\der \b r'\ 
\deltaup_n(\b r' - \b r)\, \sign(v)\,\Delta \eta \notag
\\&= \sign(v)\Delta \eta,\label{jump-deriv3}
\end{align}
where $V_{\epsilon}(\b r, t^\ast)$ corresponds to the region in space swept by $S(t)$ during $t\in[t^\ast_-, t^\ast_+]$. If $t^\ast$ is not the arrival time at $\b r$, then for sufficiently small $\epsilon$ and sufficiently large $n$, the support of $\deltaup_n(\b r' \!-\! \b r)$ is not contained in $V_\epsilon(\b r, t^\ast)$, the integral in the right hand side of Eq.~\eqref{jump-deriv3} is zero, and $\eta$ is unchanged. An alternative derivation of the jump property~\eqref{jump} based on the balance equation of the indicator function of an evolving domain is provided in Sec.~\ref{appx:indicator-gov-eq}.

\subsection{Local curvilinear partition of space}\label{appx:curvilinear-partition}
We first show that it is possible to define a local curvilinear coordinate system around the point $\b r$ with $N-1$ coordinates parallel to $S(t)$ and one coordinate perpendicular to $S(t)$ for $t$ around the arrival time $t^\ast$ at $\b r$. This local curvilinear coordinate system defines a local partition of the space around $\b r$ such that an infinitesimal volume element $\der \b r$ will be represented by $|v|\der t\der\sigma$.

Let $\b\psi(\b u, t)$ be a local parameterisation of the manifold $S(t)$ around the point $\b r$, where $\b u$ belongs to an open subset $V \subset \mathbb{R}^{N-1}$. Under appropriate regularity conditions on the normal velocity $v$ and on $S(t)$ it is always possible to choose the time dependence of $\b\psi$ such that the curves $t\mapsto \b\psi(\b u, t)$ define trajectories normal to $S(t)$ around $t^\ast$ for all $\b u$, by solving the differential equation
\begin{align}\label{orthogonal-parameterisation}
\pd{\b\psi}{t} = v(\b\psi, t)\b n(\b\psi,t)
\end{align}
from an initial parameterisation. In particular, $S(t)$ must have no `corners' in a small neighbourhood of $\b r$, $S(t)$ must be an `evolving hypersurface' around $\b r$~\cite{giga}. The parameterisation thus obtained,
\begin{align}\label{curvilinear-transform}
    \b r' = \b\psi(\b u, t) \in S(t),
\end{align}
can be seen as a coordinate transformation that maps the curvilinear coordinates $(\b u, t)$ to the cartesian coordinates $\b r'$. In doing so, time lines become replaced by the distance travelled along trajectories perpendicular to $S(t)$ and lines parameterised by $u_i$ become replaced by the distance travelled along trajectories parallel to $S(t)$. Applying the coordinate transformation~\eqref{curvilinear-transform} to an integral over space replaces the infinitesimal volume element $\der \b r'$ with
\begin{align}
    \der\b r' = |J| \der\b u \der t, 
\end{align}
where
\begin{align}
    J = \det(\der\b\psi) \equiv\det
\begin{pmatrix}
    \pd{\b\psi}{u_1}\ \cdots\ \pd{\b\psi}{u_{N-1}}\ \pd{\b\psi}{t}
\end{pmatrix}
\end{align}
is the Jacobian of the transformation~\eqref{curvilinear-transform}. The absolute value of this determinant corresponds to the volume of the $N$-dimensional parallelepiped that has the vectors $\pd{\b\psi}{u_1}, \ldots, \pd{\b\psi}{u_{N-1}}$, and $\pd{\b\psi}{t}$ as adjacent edges. This volume is equal to the volume of the $N\!-\!1$ dimensional parallelepiped defined by the vectors $\pd{\b\psi}{u_1}, \ldots, \pd{\b\psi}{u_{N-1}}$ (base area) multiplied by the projection of $\pd{\b\psi}{t}$ onto the axis perpendicular to this base (i.e., multiplied by the height)~[\citealp[Sec.~8.6.2]{cantrell}; \citealp[Sec.~2.5.4]{makarov-podkorytov}]. Because $\pd{\b\psi}{u_1}, \ldots, \pd{\b\psi}{u_{N-1}}$ all belong to the $N\!-\!1$ dimensional tangent vector space of $S(t)$ at $\b r'=\b\psi(\b u, t)$, the unit vector normal to this base is the unit normal vector $\b n$, so that the height is $\big|\pd{\b\psi}{t}\cdot\b n\big|$. Furthermore, the volume of the $N\!-\!1$ parallelepiped defined by $\pd{\b\psi}{u_1}, \ldots, \pd{\b\psi}{u_{N-1}}$ is equal to $\sqrt{\der_{\b u} \b\psi^\text{T}\, \der_{\b u}\b\psi}$, where $\der_{\b u}\b\psi$ is the $N\!\times\!(N\!-\!1)$ matrix $\begin{pmatrix}\pd{\b\psi}{u_1}\cdots \pd{\b\psi}{u_{N-1}}\end{pmatrix}$~[\citealp[Sec.~8.6.3]{cantrell}; \citealp[Sec.~2.5.3]{makarov-podkorytov}]. With Eq.~\eqref{orthogonal-parameterisation}, we thus obtain:
\begin{align*}
    |J| &= \sqrt{\der_{\b u} \b\psi^\text{T}\, \der_{\b u}\b\psi}\ \ \big|\tpd{\b\psi}{t}\cdot \b n\big| = \sqrt{\der_{\b u} \b\psi^\text{T}\, \der_{\b u}\b\psi}\ \ |v|
\end{align*}
Since the measure in surface integrals over manifolds is defined as $\der\sigma = \sqrt{\der_{\b u} \b\psi^\text{T}\, \der_{\b u}\b\psi}\ \der\b u$~\cite{makarov-podkorytov}, we finally retrieve Eq.~\eqref{curvilinear-partition}:
\begin{align}\label{curvilinear-partition-appx}
    \der \b r' = \sqrt{\der_{\b u} \b\psi^\text{T}\, \der_{\b u}\b\psi}\ \der \b u\, | v |\, \der t = \der\sigma\ | v |\, \der t
\end{align}
Note that for the transformation~\eqref{curvilinear-transform} to be injective, it is necessary that its Jacobian is nonzero, and thus that $v\neq 0$ in the neighbourhood of $\b r$, meaning that no reversal of the direction of propagation of the interface is assumed around~$\b r$.

\subsection{Representations of the surface distribution}\label{appx:surface-distrib-representations}
A similar partition of space~\eqref{curvilinear-partition-appx} can be defined in a neighbourhood of a surface $S$ with $N\!-\!1$ coordinates parallel to $S$ and one coordinate perpendicular to $S$. Let $\b r' = \b\psi(\b u, s)\in\mathbb{R}^N$ where $\b\psi(\b u, 0)$ is a parameterisation of $S$ with $\b\psi(\b 0, 0)=\b r\in S$, and with the dependence on $s$ such that
$
    \pd{\b\psi}{s} = \b n(\b\psi, s) 
$ 
in a small neighbourhood of $s=0$. The variable $s$ plays the same role as time $t$ in the developments~\eqref{orthogonal-parameterisation}--\eqref{curvilinear-partition-appx}, except that it corresponds directly to the arc length along trajectories perpendicular to $S$, i.e., $\der s$ corresponds to $|v| \der t$ in Eq.~\eqref{curvilinear-partition-appx} and we have $\der\b r' = \der\sigma\ \der s$~\cite{bedeaux-albano-mazur-1976}. This curvilinear partition of space in a small band around $S$ implies in particular that for $\b r'$ in this band:
\begin{align}\label{dirac-partitioned}
  \deltaup(\b r'-\b r) = \deltaup(s) \frac{\deltaup(\b u)}{\sqrt{\der_{\b u}\b\psi^\text{T}\, \der_{\b u}\b\psi}},
\end{align}
and
\begin{align}\label{surface-distrib-arc-length}
    \deltaup_S(\b r') = \deltaup(s).
\end{align}
Eq.~\eqref{dirac-partitioned} represents the factorisation of the Dirac distribution into the coordinates $\b u$ parrallel to $S$ and the coordinate $s$ perpendicular to $S$. The denominator accounts for the fact that $S$ is curved. If $S$ is flat and parameterised by orthonormal coordinates, the denominator is one and Eq.~\eqref{dirac-partitioned} corresponds (up to a rotation) to the well-known factorisation of the Dirac distribution in cartesian coordinates. It has to be emphasised that for Eqs~\eqref{dirac-partitioned}--\eqref{surface-distrib-arc-length} to hold, $s$ must be the arc length of a trajectory normal to $S$.

To show Eq.~\eqref{dirac-partitioned} we integrate its right hand side over space with a test function $\varphi$ and use $\der \b r' = \der\sigma\ \der s$:
\begin{align}
    &\int\hspace{-0.3em}\der\b r'\, \frac{\deltaup(s)\deltaup(\b u)}{\sqrt{\der_{\b u}\b\psi^\text{T}\, \der_{\b u}\b\psi}}\, \varphi(\b r') = \int\hspace{-0.5em}\der s\!\!\int\hspace{-0.5em}\der\sigma\, \frac{\deltaup(s) \deltaup(\b u)}{\sqrt{\der_{\b u}\b\psi^\text{T}\, \der_{\b u}\b\psi}}\,\varphi\big(\b \psi(\b u,s)\big) \notag
\\&= \int\hspace{-0.5em}\der s\int\hspace{-0.5em}\der\b u\ \deltaup(s) \deltaup(\b u)\ \varphi\big(\b \psi(\b u, s)\big) = \varphi\big(\b \psi(\b 0, 0)\big) = \varphi(\b r)
\end{align}
We proceed similarly to show Eq.~\eqref{surface-distrib-arc-length}:
\begin{align}
    &\int\hspace{-0.3em}\der\b r'\,  \deltaup(s)\, \varphi(\b r') = \int\hspace{-0.5em}\der s\!\!\int\hspace{-0.5em}\der\sigma\ \deltaup(s) \ \varphi\big(\b\psi(\b u, s)\big)\notag
\\&=\int\hspace{-0.5em}\der\sigma\ \varphi\big(\b\psi(\b u, 0)\big) = \int\hspace{-0.3em}\der \b r'\, \deltaup_S(\b r')\ \varphi(\b r')
\end{align}

Let $\Omega\subset\mathbb{R}^N$ be a domain with boundary $\p\Omega = S$, $\mathbbm{1}_\Omega$ be the indicator function of $\Omega$, and $\b n$ be the outward-pointing unit normal vector of $S$. Then
\begin{align}\label{surface-distrib-grad-indicator}
    \deltaup_S(\b r) = - \b n(\b r) \cdot\b\nabla \mathbbm{1}_\Omega(\b r).
\end{align}
Eq.~\eqref{surface-distrib-grad-indicator} was derived in~\cite{bedeaux-albano-mazur-1976,gray-lee-1977,ronis-bedeaux-oppenheim-1978,albano-bedeaux-vlieger-1979,cushman,gray1982,buenzli2015} along with the evolution equation of the indicator function of an evolving domain (see also Sec.~\ref{appx:indicator-gov-eq} below). The result~\eqref{surface-distrib-arc-length} with the identification~\eqref{surface-distrib-grad-indicator} corresponds to Eq.~(33) in Section~8.3 of Jones~\cite{jones-distrib}. Up to a more general normalisation, Eq.~\eqref{surface-distrib-arc-length} is taken as definition of $\deltaup_S$ in~\cite[Eq.~(2.11)]{ronis-bedeaux-oppenheim-1978}.

When the surface $S$ is defined implicitly as the zero level of a function $\phi(\b r)$, then $\mathbbm{1}_\Omega(\b r) = \Thetaup\big(\phi(\b r)\big)$, where $\Thetaup$ is the Heaviside step function, and one obtains from~\eqref{surface-distrib-grad-indicator} the following representation of the surface distribution:
\begin{align}\label{surface-distrib-zero-level}
    \deltaup_S(\b r) = \deltaup\big(\phi(\b r)\big) \big|\b\nabla\phi(\b r)\big|.
\end{align}
This representation of the surface distribution is taken as definition of $\deltaup_S$ in~\cite{bedeaux-albano-mazur-1976,albano-bedeaux-vlieger-1979}. It appears in some developments of the level set method~\cite{osher-fedkiw}. Note that Eq.~\eqref{surface-distrib-zero-level} with Eq.~\eqref{surface-distrib-arc-length} corresponds to Eq.~(34) in Section~8.4 of~\cite{jones-distrib}.

In~\cite{li-ito} the surface distribution appears as the kernel operator
\begin{align}\label{surface-distrib-kernel}
    \deltaup_S(\b r) = \int_S\hspace{-0.3em}\der\sigma(\b u)\ \deltaup\big(\b r - \b\psi(\b u)\big) 
\end{align}
where $\b\psi(\b r)$ is a parameterisation of $S$. Indeed, integrating the right hand side of Eq.~\eqref{surface-distrib-kernel} over space with a test function $\varphi$ gives
\begin{align}
    \int\hspace{-0.3em}\der \b r \int_S\hspace{-0.3em}\der\sigma(\b u)\ \deltaup\big(\b r - \b\psi(\b u)\big)\, \varphi(\b r) = \int_S\hspace{-0.3em}\der\sigma(\b u)\ \varphi\big(\psi(\b u)\big)
\end{align}
One may use also use Eqs~\eqref{dirac-partitioned} in Eq.~\eqref{surface-distrib-kernel} to show that it reduces to the representation~\eqref{surface-distrib-arc-length}.

Equations~\eqref{surface-distrib-arc-length}, \eqref{surface-distrib-grad-indicator}, \eqref{surface-distrib-zero-level}, and \eqref{surface-distrib-kernel} are all different representations of the surface distribution defined by~\eqref{def-surface-distrib}. These representations have been used previously in the literature, but were not necessarily identified with a distribution $\deltaup_S$ defined by~\eqref{def-surface-distrib} with the properties summarised here, with the notable exception of the early works~\cite{bedeaux-albano-mazur-1976,ronis-bedeaux-oppenheim-1978,albano-bedeaux-vlieger-1979}. Among the references cited here, Jones~\cite{jones-distrib} probably provides the most rigorous accounts on these representations based on distribution theory, however, without using the suggestive notation~$\deltaup_S(\b r)$.

\subsection{Balance equation of the indicator function of an evolving domain}\label{appx:indicator-gov-eq}
The balance equation of the indicator function of an evolving domain can be seen as a particular case of Eq.~\eqref{gov-eq} in which $\etagen = 1$ and $\eta(\b r, t) \equiv \mathbbm{1}_{\Omega(t)}(\b r) \in \{0,1\}$ jumps discontinuously between the values $0$ and $1$, such that:
\begin{align}\label{indicator-gov-eq}
    \pd{}{t}\mathbbm{1}_{\Omega(t)} = \vform\, \deltaup_{S(t)}(\b r) - \vres\, \deltaup_{S^-(t)}(\b r) = v\ \deltaup_{S(t)}(\b r) = - v\b n\cdot \b\nabla\mathbbm{1}_{\Omega(t)},
\end{align}
where the regularisation $S^-(t)$ on resorption surfaces is implicit in the last two equalities. Eq.~\eqref{indicator-gov-eq} and the representation~\eqref{surface-distrib-grad-indicator} were first derived in~\cite{bedeaux-albano-mazur-1976,ronis-bedeaux-oppenheim-1978,albano-bedeaux-vlieger-1979}. They were also derived heuristically in~\cite{gray-lee-1977,gray1982}, and proved more rigorously using distribution theory and regularised indicator functions in~\cite{cushman}. These equations were rederived in~\cite[Appendix]{buenzli2015}. The distributions defined by the gradient and Laplacian of the domain indicator function where also investigated in~\cite{bedeaux-albano-mazur-1976,ronis-bedeaux-oppenheim-1978,albano-bedeaux-vlieger-1979}, and more recently in~\cite{lange}.

The balance equation of the indicator function, Eq.~\eqref{indicator-gov-eq}, provides in fact an alternative derivation of Eq.~\eqref{jump}. Replacing $v\deltaup_{S(t)}$ by $\pd{}{t}\mathbbm{1}_{\Omega(t)}$ in Eq.~\eqref{jump-deriv1}, and integrating explicitly this sole remaining time dependence, the jump in $\eta$ at $\b r$ at the arrival time $t^\ast$ is given by
\begin{align}
    &\eta(\b r, t^\ast_+) - \eta(\b r, t^\ast_-) \notag
\\&= \lim_{\epsilon\to 0}\lim_{n\to\infty} \int\hspace{-0.5em}\der\b r'\ \deltaup_n(\b r'\!-\!\b r)\, \Delta \eta(\b r', t^\ast) \left(\mathbbm{1}_{\Omega(t^\ast_+)}(\b r') - \mathbbm{1}_{\Omega(t^\ast_-)}(\b r')\right)\notag
\\&=\Delta\eta(\b r, t^\ast)\ \lim_{\epsilon\to 0} \left(\mathbbm{1}_{\Omega(t^\ast_+)}(\b r) - \mathbbm{1}_{\Omega(t^\ast_-)}(\b r)\right)\notag
\\& = \Delta\eta(\b r, t^\ast)\ \sign\big(v(\b r, t^\ast)\big).\notag
\end{align}

\subsection{Volumetric density of a surface-bound quantity}\label{appx:volumetric-surface-density}
The surface distribution enables a simple expression for the volumetric density $n(\b r)$ of a quantity concentrated on a surface $S$:
\begin{align}\label{surface-to-volume-density}
    n(\b r) = \rho(\b r)\ \deltaup_S(\b r),
\end{align}
where $\rho$ is the quantity's \emph{surface density} on $S$, and $\deltaup_S(\b r)$ is the surface distribution defined by Eq.~\eqref{def-surface-distrib}. Eq.~\eqref{surface-to-volume-density} can be shown by integrating it over a neighbourhood $V\subset\mathbb{R}^N$ of a point $\b r$ on the surface $S$. The left hand side gives, by definition of $n$, the absolute amount of the quantity found in the volume $V$. With the definition~\eqref{def-surface-distrib}, the right hand side gives $\int_{V\cap S}\der \sigma(\b p)\ \rho(\b p)$, which by definition of $\rho$ is also the absolute amount of the quantity found on $S$ in the volume $V$.

Alternatively, the volumetric density of point particles $i$ in space is
\begin{align}\label{density-point-particles}
n(\b r) = \sum_i\ \deltaup(\b r - \b r_i).
\end{align}
Assuming the particles all belong to $S$ and using the partition of space~\eqref{dirac-partitioned} and the representation~\eqref{surface-distrib-arc-length}, one has
\begin{align}
    n(\b r) = \deltaup(s) \sum_i\frac{\deltaup(\b u - \b u_i)}{\sqrt{\der_{\b u}\b\psi^\text{T}\ \der_{\b u}\b\psi}} = \deltaup_S(\b r)\ \rho(\b r)
\end{align}
where the surface density on the curved manifold~$S$ parameterised by $\b\psi(\b u)$ is represented as
\begin{align}
\rho(\b r) \equiv \sum_i \frac{\deltaup(\b u - \b u_i)}{\sqrt{\der_{\b u}\b\psi^\text{T}\ \der_{\b u}\b\psi}}.
\end{align}

\section{Numerical discretisation}\label{appx:numerical}
The governing equations for $\eta(\b r, t)$ in the one-dimensional and two-dimensional examples were solved numerically based on a simple explicit scheme, using forward finite difference in time (Euler) and a fixed discretisation grid of the computational domain $V$. The singular surface terms were implemented by explicitly tracking the position of the interface and enforcing the jump condition~\eqref{jump} at this interface. The following steps were performed for each time increment~$\Delta t$:
\begin{enumerate}
    \item Evolve the interface given $v$. Determine the set of discretisation points $V_\form$ at which $\eta$ was formed and the set of discretisation points $V_\res$ at which $\eta$ was resorbed;
    \item For each point $\b r_i \in V_\form$, increase $\eta$ by $\eta_\form(\b r_i, t) - \eta(\b r_i, t)$;
    \item For each point $\b r_i \in V_\res$, set $\eta$ to $0$;
    \item For each point $\b r_i \in V$, add to $\eta$ the quantity $\Delta t \mathcal{F}(\eta(\b r_i, t))$.
\end{enumerate}
In point 2., $\eta(\b r_i, t)$ is substracted so that the value $\eta_\form$ is generated even if $\eta$ has a residual value at $\b r_i$. This can happen at reversal points between resorption and formation due to round-off errors.

\begin{figure}
    \centering\includegraphics[width=\figurewidth]{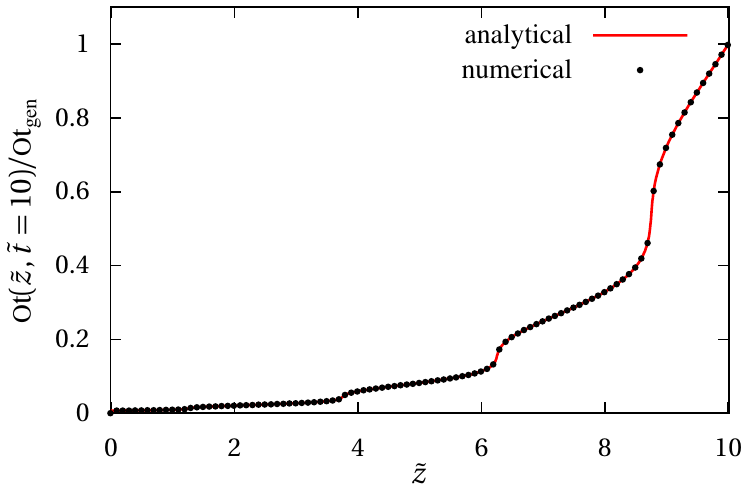}
    \caption{Comparison between numerical solution obtained by integrating Eq.~\eqref{ocy} (dots) and (semi-)analytical solution provided by Eq.~\eqref{ocy-theor} (solid red line) at time $\tilde{t}=10$.}\label{fig:num-analytic}
\end{figure}

In Figure~\ref{fig:num-analytic}, we compare a direct simulation of Eq.~\eqref{ocy} with the analytical result~\eqref{ocy-theor} in the same one-dimensional situation as Figure~\ref{fig:ocy}a. The analytic solution~\eqref{ocy-theor} requires the arrival time $\tilde{t}^\ast(\tilde{z})$ (in dimensionless coordinate), i.e. the time at which the interface $S(\tilde{t})$ reaches the point $\tilde{z}$. In the situation depicted in Figure~\ref{fig:num-analytic}, the arrival time is found numerically by solving $\tilde{z} = S(\tilde{t}^\ast)$ for $\tilde{t}^\ast$ using Newton's method, where $S(\tilde{t}) = \tilde{t} + \alpha \sin\left(\frac{2\pi n \tilde{t}}{\tilde{t}_f}\right)$ with $\alpha=0.35, n=4, \tilde{t}_f=10$.

\pbuenzli{The double remodelling events simulated in Figure~\ref{fig:miner2} and Figure~\ref{fig:bmd-averages} assumed given populations of osteoblasts and osteoclasts $\rho_\ob$, $\rho_\oc$ and constant secretory and resorption rates $\kform$, $\kres$, such that the normal velocity of the interface $v=\kform\,\ob>0$ in formation, $v=\kres\,\oc<0$ in resorption, was given by:}
\begin{align}
    \pbuenzli{v(x,t) = \begin{cases}v_0 &\quad 0 < t \leq t_0,
    \\v_1 \sin\left(\pi\frac{x-a_1}{b_1-a_1}\right)\cos\left(\pi\frac{t-t_1^\text{beg}}{t_1^\text{end}-t_1^\text{beg}}\right)& \quad t_1^\text{beg} < t \leq t_1^\text{end},
    \\v_2 \sin\left(\pi\frac{x-a_2}{b_2-a_2}\right)\cos\left(\pi\frac{t-t_2^\text{beg}}{t_2^\text{end}-t_2^\text{beg}}\right)& \quad t_2^\text{beg} < t \leq t_2^\text{end},
    \\0 & \quad \text{otherwise}
    \end{cases}}
\end{align}
\pbuenzli{where $v_0=3\,\um/\da$, $t_0=25\,\days$; $v_1=0.62\,\um/\da$, $t_1^\text{beg}=9.5\,\yr$, $t_1^\text{end}=10\,\yr$, $a_1=25\,\um$, $b_1=275\,\um$; $v_2=0.49\,\um/\da$, $t_2^\text{beg}=19.5\,\yr$, $t_2^\text{end}=20\,\yr$, $a_2=75\,\um$, $b_2=425\,\um$. Times $0<t\leq t_0$ correspond to a phase of bone tissue growth, times $t_1^\text{beg} < t \leq t_1^\text{end}$ to the first remodelling event, and times $t_2^\text{beg} < t \leq t_2^\text{end}$ to the second remodelling event.}

\end{appendices}

\end{document}